\documentstyle[emulapj]{article}

\begin{document}

\lefthead{MAGNETO-CENTRIFUGALLY DRIVEN WINDS}
%:
%COMPARISON OF MHD SIMULATIONS
%WITH THEORY}
\righthead{USTYUGOVA ET AL.}

\accepted{}

\title{Magneto-centrifugally Driven Winds:\\
Comparison of MHD Simulations with Theory}

\medskip

\author{G.V. Ustyugova}
\affil{Keldysh Institute of
Applied Mathematics, Russian Academy
of Sciences, Moscow, Russia, 125047,
ustyugg@spp.Keldysh.ru}
\author{A.V. Koldoba}
\affil{Institute of
Mathematical Modelling,
Russian Academy of Sciences, Moscow,
Russia, 125047}
\author{M.M. Romanova}
\affil{Space Research Institute,
Russian Academy of Sciences, Moscow, Russia; and\\
Department of Astronomy,
Cornell University, Ithaca, NY 14853-6801;
romanova@astrosun.tn.cornell.edu}
\author{V.M. Chechetkin}
\affil{Keldysh Institute of
Applied Mathematics, Russian Academy
of Sciences, Moscow, Russia;
chech@spp.Keldysh.ru}
\author{R.V.E. Lovelace}
\affil{Department of Astronomy,
Cornell University, Ithaca, NY 14853-6801;
rvl1@cornell.edu }

\slugcomment{Accepted to the Astrophysical Journal}

\begin{abstract}

 Stationary magnetohydrodynamic (MHD)
outflows from a rotating
accretion disk are investigated numerically
by time-dependent axisymmetric
simulations.
 The initial magnetic field is taken to
be a split-monopole poloidal field configuration
frozen into the
disk.
 The disk is treated as a perfectly
conducting, time-independent density
 boundary [$\rho(r)$]
in Keplerian rotation.
  The outflow velocity
from this surface is not specified but rather
is determined self-consistently from the MHD
equations.
   The temperature of
the matter outflowing from
the disk is  small in the
region  where the magnetic
field is inclined away from
the symmetry axis
($c_s^2 \ll v_K^2$), but
relatively high
($c_s^2\lesssim v_K^2$) at very small radii in the
disk
where the magnetic field is not inclined
away from the axis.
   We have found a large class of stationary
MHD winds.
   Within the simulation region, the outflow
accelerates from thermal velocity ($\sim c_s$)
to a much larger asymptotic poloidal flow
velocity of the order of one-half $\sqrt{GM/r_i}$
where $M$ is the mass of the central object
and $r_i$ is the inner radius of the disk.
   This asymptotic velocity is much larger than
the local escape speed and is larger than
fast magnetosonic speed by a factor of $\sim 1.75$.
  The {\it acceleration distance} for the outflow, over
which the flow accelerates from $\sim 0$
to, say, $90\%$ of the asymptotic speed, occurs
at a flow distance of about $80 r_i$.
   The outflows are approximately spherical, with
only small collimation within the simulation region.
   The {\it collimation distance} over which
the flow becomes collimated (with divergence less
than, say, $10^o$) is much larger than the size of
our simulation region.
   Close to the disk the outflow is driven by
the centrifugal force while at all larger
distances the flow is driven by the
magnetic force which is proportional to $-{\bf \nabla}
(rB_\phi)^2$, where $B_\phi$ is the toroidal field.

  Our stationary numerical solutions allow us
(1) to compare the results  with MHD theory
of stationary flows,
(2) to
investigate the influence of different
outer boundary conditions
on the flows, and
(3) to investigate the influence
of the shape of the simulation region
 on the flows.
   Different comparisons were made with the
theory.
   The ideal MHD integrals of motion (constants
on flux surfaces) were
calculated along magnetic field
lines and were shown
to be constants with accuracy $5-15 \%$.
   Other characteristics of the
numerical solutions were compared with the theory,
including conditions at the
Alfv\'en surface.

    Different outer boundary conditions
on the  toroidal component
of the magnetic field were investigated.
    We conclude that
the commonly used ``free'' boundary condition on
the toroidal field leads
to  artificial magnetic
forces on the outer boundaries,
which can significantly influence to the
calculated flows.
   New outer boundary conditions
are proposed and investigated which do not
give artifical forces.

   We show that simulated flows
may depend  on the shape
of the simulation region.
   Namely, if the simulation region
is elongated in the $z-$direction,
then Mach cones on the outer
cylindrical boundary may be
partially directed inside the
simulation region.
   Because of this, the boundary can have
an artificial influence on the calculated
flow.
   This effect is reduced if the
computational region is approximately square
or if it is spherical.
  Simulations of MHD outflows with an
elongated computational
region can lead to
{\it artificial} collimation
of the flow.

\end {abstract}


\section{Introduction}

   The  existence and nature of
magnetohydrodynamic (MHD)
outflows from an accretion disk
threaded by an ordered magnetic
field is a long-standing  astrophysical problem.
   The problem has been investigated
theoretically by many
authors
 (Blandford \& Payne 1982;
Pudritz \& Norman 1986; Sakurai 1987;
Koupelis \& Van Horn 1989; Lovelace, Berk \&
 Contopoulos 1991;  Pelletier \& Pudritz 1992;
K\"onigl \& Ruden 1993;
Cao \& Spruit 1994;
 Contopoulos \& Lovelace 1994;
Contopoulos 1995; Ostriker 1997).
See also reviews by Bisnovatyi-Kogan (1993)
and Livio (1997).
 From the theory,
a necessary condition for
magnetically/centrifugally
driven outflows  is that
the poloidal magnetic field at the
disk's surface be inclined away from the
symmetry axis ($z$) at a sufficiently large
angle.

   However, the analytical theory
makes  drastic simplifications
such as assuming self-similar dependences
on the radial distance ($r$ in cylindrical
coordinates),
or by integrating over the cross-section
of the outflow.  The self-similar
solutions have divergences at both
small and large $r$ so that the influence
of these regions is unknown.

   Numerical MHD simulations are
essential to establish
the existence and understand the
nature of
magnetically/centrifugally
driven outflows.
      Stationary and non-stationary MHD flows were
investigated by Kudoh \& Shibata
(1995, 1997a,b) in one-dimensional ($1.5$D)
simulations.
   These simulations allowed an
investigation of outflows
for a wide range of parameters.
  However, they supposed a fixed configuration
of the poloidal magnetic field.
   Two-dimensional ($2.5$D) simulations
of outflows from accretion disks were
performed by
Uchida \& Shibata (1985),
Shibata \& Uchida (1986),
Stone \& Norman (1994),
Matsumoto et al. (1996).
    These simulations led to strongly non-stationary
accretion and outflows from the disk.
  In most of these studies, the
non-stationarity of the solutions is
due to the start up conditions with
the disk rotating but the corona
of the disk not rotating.
   In other cases the non-stationarity is
due to the disk rotating at a
significantly sub-Keplerian rate.
   These simulations are valuable in showing
that temporary MHD outflows
are possible, but the results depend
strongly on the assumed initial conditions.

   In order to
avoid the strong dependence
on initial conditions and the
problems associated with following
the internal dynamics of the accretion
disk, we earlier proposed treating the outer,
surface layers of the disk as a boundary condition
(Ustyugova et al. 1995; Koldoba et al. 1996;
Romanova et al. 1997; Romanova et al. 1998).
  This approach has
been followed by others (Ouyed \& Pudritz 1997;
Ouyed, Pudritz \& Stone 1997;
Meier et al. 1997).
   In these simulations
the ``disk'' represents an
outer layer of the accretion
disk.
    In actual situation, the outflowing matter
will affect the disk evolution, or at least
to the evolution of the
surface layers of the disk.
  The angular momentum carried
away by MHD outflows can
give a disk accretion
rate much larger than
the viscous accretion rate
of say an $\alpha$-disk,
  but the accretion speeds
are typically much smaller
than the free-fall speed
(Lovelace, Romanova \& Newman
1994; Lovelace, Newman \& Romanova 1997).
   Thus, the disk can be
treated as stationary
during the formation and establishment
of MHD outflows which takes
place on a free-fall time scale.
   However, the long-time simulations of outflows
including the back reaction on the disk
are clearly of interest for future research.

   Different initial magnetic
field configurations
have been assumed in earlier studies.
   The initial field
assumed by Ouyed \& Pudritz (1997) was the Cao
\& Spruit (1994) field
 which decreases
slowly with radial distance
on the disk's surface.
   On the other hand, the initial magnetic
field of Ustyugova et al. (1995)
was the split-monopole field
 (Sakurai 1978; 1985),
which decreases rapidly with radial
distance on the disk surface.
The temperature of matter outflowing
from the disk of Ouyed \& Pudritz (1997)
was small, and the initial
 magnetic field was weak.
   However, Ouyed \& Pudritz (1997)
introduced a spectrum of
turbulent Alfv\'en
waves with a high pressure which is
similar to having a high temperature corona.
    Thus the approach of
 Ouyed \& Pudritz (1997) is similar to that
of Ustyugova et al. (1995) where the magnetic
field is weak and the coronal temperature
is high.
   In both papers, the initial twist of
the magnetic field results from the disk
rotation because the corona is not rotating.
   This twisting of the magnetic
field gives the collimation observed
in both papers.

   It is important to get stationary outflows
using time-dependent MHD equations because
the non-stationary flows may be artifacts
of the initial conditions.
   Stationary magneto-centrifugally driven
outflows for relatively low temperature of
the ``disk'' matter were obtained
in the $2.5$D simulations by Romanova et al. (1997)
for the case where the initial magnetic
field was a ``tapered'' split monopole type field.
  This work found that in the stationary state
the outflow was quasi-spherical with
essentially no collimation within the simulation
region.
 Close to the disk the outflow
was driven by the centrifugal
force while at larger distances the
magnetic force was dominant.

   In this work we investigate the case of a pure
(that is, non-tapered) split-monopole magnetic field
 by axisymmetric (2.5D) numerical simulations.
The motivation was to study MHD outflows from
a relatively cold accretion disk where
magnetic field lines are  inclined away
from the symmetry axis.
   To remove the influence
of the region near the axis where magnetic
field lines are not significantly inclined, we pushed hot
matter from the disk in the small area around the axis.
   We compare our
simulation results  with the theory
of stationary MHD flows.
   Further, we use our stationary
simulation flows
to investigate the influence of
outer boundary conditions.
   Our earlier study
(Romanova et al. 1997)
showed that some simple outer boundary
 conditions on the
toroidal magnetic field
can lead to artificial
forces on the boundary which
significantly influence the flow within
the simulation region.
Here, we consider in further detail the
influence of outer boundary conditions on
the calculated flows.

   In \S 2 the theory of stationary MHD
flows is briefly reviewed.
   In \S 3 the numerical model is presented.
 The influence of the outer boundary condition
on the toroidal magnetic field and
the shape of the computational
region is analyzed in \S 4.
   In \S 5 we present results
of simulations of stationary flows and compare
them with theory.
   In \S 6
conclusions of this work are summarized.

\section{Theory of Stationary MHD Flows}

  The theory of stationary,
axisymmetric, ideal MHD flows
was developed by Chandrasekhar (1956),
Woltjer (1959), Mestel (1961),
Kulikovskyi \& Lyubimov (1962), and others.
   Under these conditions the
MHD equations can be reduced to
a single equation for the ``flux function''
$\Psi(r,z)$ in cylindrical $(r,\phi,z)$
coordinates (Heinemann \& Olbert 1978; Lovelace
et al. 1986).
   The flux function
$\Psi$ labels flux surfaces so that
$\Psi(r,z)=$const represents the poloidal
projection of a field line.
  The equation for $\Psi$ is commonly referred to as
the Grad-Shafranov equation (Lovelace et al. 1986).

\subsection{Integrals of Motion}

 For axisymmetric conditions the flow field
can be written as ${\bf v}={\bf v}_p
+ v_\phi {\bf e}_\phi$ where ${\bf v}_p$
is the poloidal $(r,z)$ component, $v_\phi =
\omega r >0$ is the toroidal component, and
${\bf e}_\phi$ is the unit toroidal vector.
Similarly, the magnetic field can be
written as ${\bf B} =
{\bf B}_p + B_\phi {\bf e}_\phi$.
The ideal MHD equations then imply
that certain quantities are constants
on any given flux surface $\Psi(r,z)=$const
or equivalently they are constants
along any given stream line or
a given magnetic field line.
These integrals are  functions of
$\Psi$
(see for example Lovelace et al. 1986),
$$
{\bf v}_p  =
\frac{K(\Psi)}{4 \pi \rho} {\bf
B}_p
\eqno(1)
$$
$$
 \omega r^2 - \frac{r B_\phi}{K} =
\Lambda(\Psi)
\eqno(2)
$$
$$
\omega - \frac{K B_\phi}{4 \pi
\rho r} = \Omega(\Psi)
\eqno(3)
$$
$$
S = S(\Psi)
\eqno(4)
$$
$$
 w + \Phi - \frac{\Omega^2 r^2}{2} +
\frac{{\bf v}_p^2}{2}  + \frac{1}{2}
( \omega - \Omega )^2 r^2 = E(\Psi)
\eqno(5)
$$
Here, $S$ is the entropy,
$w$ is the enthalpy, and
$\Phi$ is the gravitational
potential. The quantity $K$
corresponds to the conservation
of mass along a streamline,
$\Lambda$ to the conservation
of angular momentum, $\Omega$
to the conservation of helicity, $S$
to the conservation of entropy,
and $E$ (Bernoulli's constant) to
the conservation of energy.

  The remaining MHD equation (which cannot
be written in the integral form)
is the Euler force equation across
the poloidal magnetic
 field line (Bogovalov 1997),
$$
({\bf v}_p^2 - v_{Ap}^2) \frac{\partial\theta}{\partial s}
 - \frac{\cos \theta}{r}
(v^2_\phi - v^2_{A\phi}) +\quad\quad\quad\quad
$$
$$\quad\quad\quad\quad+ \frac{1}{\rho}
\frac{\partial}{\partial n}
\left( p + \frac{{\bf B}^2}{8 \pi} \right)
+ \frac{\partial \Phi}{\partial n}  =0~,
\eqno(6)
$$
which is equivalent to the Grad-Shafranov equation.
Here,
${\partial}/{\partial n}$ is the
derivative in the direction perpendicular
to magnetic field lines
and directed outward from the axis,
$\theta$ is the angle of
inclination of the poloidal magnetic
field line away from the $z-$axis,
$s$  is the distance from the disk
along a magnetic field
line, $v_{Ap}\equiv
|{\bf B}_p|/\sqrt{4\pi\rho}$ and
$v_{A\phi}\equiv |B_\phi|/\sqrt{4\pi \rho}$
 are the poloidal and azimuthal
Alfv\'en velocities.
The quantity
${\partial\theta}/{\partial s}$ is the curvature
of magnetic field line.
The first two terms in equation (6) are
determined by the non-diagonal
(tension) part of the stress tensor,
 $\rho v_i v_k  +
\left( p + {{\bf B}^2}/{8 \pi} \right) \delta_{ik}
- {B_i B_k}/{4 \pi}$.
  The third term
is determined by the total
(matter plus magnetic) pressure
$p + {{\bf B}^2}/{8 \pi}$ and the gravity
force $\partial \Phi/\partial n$.

\subsection{Physical Sense of Integrals of Motion}

   To clarify the physical sense
of the integrals of motion,
it is useful to derive
the fluxes of mass,
angular momentum (about the $z-$axis), and energy.
   The corresponding conservation
laws for stationary conditions are
$$  {\bf\nabla}\cdot (\rho {\bf v}_p)  = 0~,
\eqno(7)
$$
$${\bf\nabla}\cdot \left (r\rho {\bf v}_p v_\phi   -
{{\bf B}_pr B_\phi  \over 4\pi}\right ) = 0 ~,
\eqno(8)$$
$${ \bf\nabla}\cdot \left [ \rho {\bf v}_p
\left ( \frac{{\bf v}^2}{2} +
\frac{{\bf B}^2}{4 \pi \rho} + w + \Phi \right )
-{{\bf B}_p({\bf v}\cdot {\bf B})\over 4\pi}
\right ] = 0 ~.
\eqno(9)
$$
  Because ${\bf v}_p \parallel {\bf B}_p$,
the vector flux densities
are directed along the field lines.

   Consider the fluxes through an annular
region with surface
area element
$d{\bf S}$.
   The matter flux through the
axisymmetric surface $\bf S$
extending out from the $z-$axis is
$${\cal F}_M  = \int \limits_{\bf S}
 d{\bf S}\cdot\rho {\bf
v}_p  = {1\over 4\pi}\int \limits_{\bf S}
d{\bf S}\cdot {\bf B}_p {K(\Psi) } ~,
\eqno(10)
 $$
where we  took into account
the integral (1).
   $d{\bf S}\cdot{\bf B}_p$ is  the magnetic flux
through the annular region bounded by flux surfaces
$\Psi$ and $\Psi + d\Psi$.
   Thus we can change from space integration
to integration over $\Psi$.
   Because
 $B_r=-(1/r){\partial\Psi/\partial z}$
and
 $B_z=(1/r){\partial \Psi/\partial r}$, we
have
$${\cal F}_M(\Psi) = \frac{1}{2}
\int \limits_0^\Psi d \Psi^\prime K(\Psi^\prime) ~,
\eqno(11)
$$
where $\Psi=0$ corresponds to the $z-$axis.
Similarly,
$${\cal F}_L(\Psi) =\int d {\bf S} \cdot
\left (r \rho {\bf v}_p
v_\phi  -
\frac{{\bf B}_p rB_\phi}{4 \pi}  \right )
$$
$$
= \frac{1}{2}
\int \limits_0^\Psi d \Psi^\prime
\Lambda (\Psi^\prime) K(\Psi^\prime)~,
\eqno(12)
$$
$${\cal F}_E(\Psi) \!=\!\! \int \!\! d{\bf S}
\cdot \rho {\bf v}_p
\left [ \left (
\frac{{\bf v}^2}{2} +
\frac{{\bf B}^2}{4 \pi \rho} + w +
\Phi \right )
 -
 \frac{ {\bf B}_p{\bf v} \cdot {\bf B}}
{4 \pi} \right ]
$$
$$=
\frac{1}{2} \int \limits_0^\Psi
d\Psi^\prime (E + \Lambda
\Omega) K~.
\eqno(13)
$$
Thus,
$Kd\Psi/2 $  is the matter
flux between the flux
surfaces separated by $d\Psi$,
$\Lambda Kd\Psi/2 $ is the angular
momentum flux,
and $(E + \Lambda \Omega) Kd\Psi/2$  is the
energy flux.
Note that
$\Lambda(\Psi)$ is specific
angular momentum carried
along the magnetic field line
$\Psi={\rm const}$,
$E(\Psi) + \Lambda \Omega (\Psi)$ is the specific energy,
and $\Omega(\Psi)$ is the angular velocity
of the disk at the point where
the magnetic field line  or
flux surface $\Psi ={\rm const}$
intersects the disk
(for $|{\bf v}_p| \rightarrow 0$ at
the disk).

\subsection{Conditions at the Alfv\'en Surface}

   Conditions at the Alfv\'en surface
are known to be important for the
global properties of MHD flows (Weber \&
Davis 1967).
  Equations (2) and
(3) constitute a  linear
system of equations
for $\omega$ and $B_\phi$.
The determinant of this system is zero if
$K^2=4 \pi \rho$.  Under this condition
  a solution exists  if
$\Lambda= r^2 \Omega$ (Weber \& Davis 1967)
which corresponds to
${\bf v}_p={\bf B}_p/{\sqrt{4 \pi \rho}}=
{\bf v}_{Ap}$.   This is the condition
which defines the Alfv\'en surface.
  Figure 1 shows a possible field line $\Psi={\rm const}$
and  Alfv\'en surface $A$.
  The radius at which this field line intersects the
disk is $r_d(\Psi)$. The radius at which it crosses
the Alfv\'en surface is $r_A(\Psi)$.
 The density
at this point on the Alfv\'en surface
is $\rho_A (\Psi)$.
Thus,

\begin{figure*}[t]
\epsscale{0.4}
\plotone{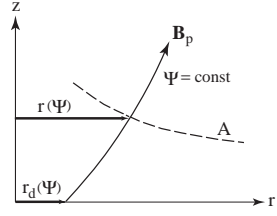}
\caption{
The figure shows
a poloidal magnetic field line $\Psi=$ const
(solid line) which starts at the radial
distance
$r=r_d(\Psi)$ on the disk and crosses the
Alfv\'en surface $A$ (dashed line) at
the radial distance $r_A(\Psi)$ from the axis.
}
\label{Figure 1}
\end{figure*}

$$
\rho_A (\Psi)= K^2 (\Psi)/ 4 \pi ~, \quad
r^2_A (\Psi)= \Lambda (\Psi) / \Omega (\Psi)~.
\eqno(14)
$$
Equations (2) and (3) give
$$
\omega = \Omega~
\frac{1-\rho_A r^2_A / {\rho r^2}}{1 -
\rho_A /\rho}~,
\eqno(15)
$$
$$
B_\phi =
r\Omega  \sqrt{4\pi \rho_A}~ \frac{1 -
 r^2_A /r^2}{1-\rho_A /\rho}.
\eqno(16)
$$
Taking into account (14) - (16),
one can express the
fluxes of mass, angular momentum and energy,
using only the values of
physical quantities on the
Alfv\`en surface:
$$
{\cal F}_M(\Psi) = \int \limits_0^\Psi d \Psi^\prime
\sqrt{\pi \rho_A}~,
\eqno(17)
$$
$$
{\cal F}_L(\Psi) = \int \limits_0^\Psi d \Psi^\prime
 \sqrt{\pi \rho_A}~\Omega ~r^2_A ~,
\eqno(18)
$$
$$
{\cal F}_E(\Psi) = \int \limits_0^\Psi
d \Psi^\prime \sqrt{\pi\rho_A}~
 \left( E+ \Omega^2 r^2_A \right)~ .
\eqno(19)
$$

\subsection{Forces}

   For understanding
the plasma acceleration, we
project the different forces
onto the poloidal magnetic
field lines.
   As mentioned,  in a stationary
state,  matter flows
along the poloidal magnetic field lines.
   The acceleration in the  poloidal
$(r,z)$ plane is
$$
({\bf v}_p \cdot {\bf \nabla}) {\bf v}_p +
v_\phi ({\bf e}_\phi \cdot {\bf \nabla} )
(v_\phi {\bf e}_\phi)~.
\eqno(20)$$
  The last term represents the centrifugal
acceleration
$ - (v^2_\phi /r ){\bf e}_r=-r\omega^2  {\bf
 e}_r$.
   To get the force per unit
mass along a magnetic field line,
we multiply the Euler equation by a
 unit vector  $\hat{\bf b}$ parallel to
${\bf B}_p$.
  This gives
$$
f = \omega^2 r \sin \theta -
\frac{1}{\rho} \frac{ \partial p}{ \partial s}
- \frac{\partial \Phi}{\partial s} + \frac{1}{4 \pi \rho}
{\hat{\bf b}} \cdot [({\bf \nabla}
\times {\bf B})\times {\bf B}]~,
\eqno(21)
$$
where $\theta$ is the inclination angle
of the field line to the $z-$axis.
The final
term of (21) is the projection
of the magnetic force in the
direction of $\hat{\bf b}$,
which can be transformed to
$$
f_M=\frac{1}{4 \pi \rho}
{\hat{\bf b}} \cdot [({\bf \nabla}
\times {\bf B})\times {\bf B}]=
-{1\over{8 \pi \rho r^2}}
{\partial(rB_\phi)^2\over\partial s}~,
$$
which is useful for
understanding our results.

  When magnetic field lines
are inclined outward, away from the
symmetry axis, the gravitational force
$f_G=\partial \Phi/\partial s$
opposes the outflow of matter
from the disk.
    If the matter is relatively cold then
the pressure gradient force
$f_P=-({1}/{\rho}) ({ \partial p}/{ \partial s})$
is unimportant.
   Then matter can be accelerated outward
 by the centrifugal force
$f_C=\omega^2 r {\rm sin} \theta$ and/or
the magnetic force $f_M$.
 This determines
the driving mechanisms of the outflow, centrifugal
and/or magnetic.
   The centrifugal force always acts to
accelerates matter outward if the distance between
magnetic field line and the axis increases.
  Consider the direction of
the magnetic force.
   Note that the lines on which
$rB_\phi=$const
are also poloidal current-density lines;  that is,
${\bf j}_p = -({\bf e}_\phi/r) \times
{\bf \nabla}(r B_\phi)$ so that
${\bf j}_p \cdot {\bf \nabla}(rB_\phi)=0$.
 Consider a configuration of
magnetic field line
$\Psi=$const and a line
 of  current-density ${\bf j}_p$ as
shown in  Figure 2.
 The poloidal component of the  magnetic
force $\propto -\nabla(rB_\phi)^2$
is perpendicular to the
current-density ${\bf j}_p$  and
is shown on Figure 2 by arrows.
   Projection of this force onto
the poloidal  magnetic field
shows that the force
pushes matter upward near the disk
(lower part of the region),
and pushes matter downward further from the disk
(upper part of the region).
   The $\phi-$component of the magnetic force
 $\propto ~{\bf j}_p\times{\bf B}_p $ acts
in the direction of the disk
rotation and leads to winding of the
magnetic field line close to the disk
and leads to unwinding of magnetic field
line farther from the disk.

%\placefigure{fig2}

\begin{figure*}[t]
\epsscale{0.4}
\plotone{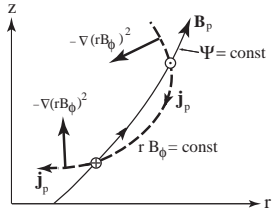}
\caption{
The sketch shows
the directions of
the magnetic forces for different configurations
of the poloidal magnetic field
${\bf B}_p$ (bold line) and the
poloidal current-density ${\bf j}_p$ (thin line).
     The magnetic force
in the poloidal plane $\propto -\nabla(r B_\phi)^2$
is perpendicular to the poloidal current-density
(bold arrows).
   The lower part of the figure shows the
case where the magnetic force  pushes matter
away from the disk,
while the top part of the figure shows
the opposite situation.
   The force in the azimuthal direction
${\bf j}_p\times{\bf B}_p/c$
acts in the direction of rotation of
the disk in the bottom part of the figure
line and in opposite direction in the top part.
}
\label{Figure 2}
\end{figure*}

  Thus, magnetic and centrifugal forces may
both accelerate matter, but this
depends on the
configuration
of magnetic field and current-density lines.

\subsection{Collimation}

   Consider now the collimation of
the flow.
   From equation (6), taking into
account that
${\rm cos} \theta = {\partial r}/{\partial n}$,
we have
$$
({\bf v}^2_p - {\ v}^2_{Ap})
\frac{\partial \theta}{\partial s} =
- \frac{1}{8 \pi \rho r^2}
\frac{\partial}{\partial n}
\left( r B_\phi \right)^2 +
\frac{\cos \theta v^2_\phi}{r}
$$
$$
- \frac{1}{\rho}
\frac{\partial}{\partial n} \left(p +
 \frac{{\bf B}^2_p}{8 \pi}
\right) - \frac{\partial \Phi}{\partial n}~.
\eqno(22)
$$
   At large distances from the Alfv\'en surface
$r\gg r_A$, the density $\rho \ll \rho_A$,
but values $\rho r^2$ and ${\bf v}_p^2$ remain
finite (Heyvaerts and Norman 1989).
   Then $v^2_{Ap}=\left(\rho/\rho_A\right)^2
{\bf v}_p^2 \ll {\bf v}_p^2$,
so that the second term on the left-hand side
of (22)
is negligible.
    On the right-hand side,
only the first term is important
for $r \gg r_A$.
    Then, equation (22) simplifies to
$$
v^2_p ~\frac{\partial \theta}{\partial s} =
- \frac{1}{8 \pi \rho r^2}
\frac{\partial}{\partial n} \left( r B_\phi \right)^2~ .
\eqno(23)
$$
   Thus, the curvature of magnetic field lines
in the region $r \gg r_A$ is
determined by the gradient
$(rB_\phi)^2$.
   Figure 3 shows examples
where two magnetic field
lines $\Psi_1$ and $\Psi_2$ cross the
line $rB_\phi=$ const.
   Again, the magnetic force
$\propto-\nabla(r B_\phi)^2$
acts in the direction
perpendicular to the current-density
${\bf j}_p$.
    Here, we are interested in the
projection of this force onto
a poloidal magnetic field
line.
   From the figure one can see
that the magnetic force acts to
``collimate'' the magnetic
field line  $\Psi_1$
and ``anticollimate'' the field
line $\Psi_2$.

\placefigure{fig3}

\begin{figure*}[t]
\epsscale{0.4}
\plotone{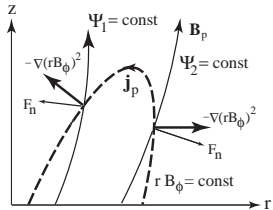}
\caption{
The figure shows a poloidal current-density
line ${\bf j}_p$ on which
$r B_\phi=$ const and two poloidal
field lines.
On the  field line
$\Psi_1$, the magnetic force $\propto
-{\bf \nabla}(rB_\phi)^2$ acts to
give collimation while for  the
field line $\Psi_2$  the force
acts to ``anti-collimate'' the flow.
}
\label{Figure 3}
\end{figure*}

\section{Numerical Simulations of
MHD Outflows}

  For our time-dependent
simulations of axisymmetric flows
of an ideal plasma in a
gravitational field
the equations are
$$
{\frac{\partial \rho}{\partial  t}}  +
{ {\bf \nabla}\cdot (\rho {\bf v})} =  0~,
\eqno(24)
$$
$$
{\frac{ \partial (\rho {\bf v})}{\partial  t}}  +
{{\bf \nabla}\cdot {\cal T}}  =  \rho {\bf g}~,
\eqno(25)
$$
$$
{\frac{ \partial {\bf B}}{\partial  t}} -
{{\bf\nabla}\times ({\bf v}\times {\bf B})}  = 0~,
\eqno(26)
$$
$$
{\frac{ \partial( \rho S)}{\partial  t}} +
{{\bf \nabla}\cdot (\rho  {\bf v} S)}  =  0~.
\eqno(27)
$$
Here, $S$ is entropy,
 ${\cal T}_{jk} = \rho v_j v_k
+ p \delta_{jk} +  ( {\bf B}^2
\delta_{jk}/2 -$ $B_j B_k)/(4\pi)$;
is the stress tensor;
${\bf g} = - {\bf \nabla} \Phi$
is the gravitational acceleration;
and
$\Phi$ is the gravitational
potential of the central object.

  The energy equation (27) is
written in conservative form.
  From (24) and (27), one also has
$$
\frac{\partial [\rho f(S)]}{\partial {\rm t}} +
{\bf {\bf \nabla}}\cdot \left [\rho f(S) {\bf
v}\right ] = 0~,
\eqno(28)
$$
for any continuous function $f(S)$.
We take the equation of state to be
$p = (\gamma -1) \rho e$,
where $e$ is specific internal energy,
and $\gamma =$ const.
In the present work $\gamma = 5/3$.
  We take
$f(S)=p/\rho^\gamma$, because the right-hand
side depends only on the entropy.
We solve the system of equations
(24)-(26) and (28) numerically.

   The central object of mass
$M$ is  at the center of
our coordinate system.
  The disk is located at $z=0$
and  is treated as a
perfectly conducting surface
rotating with Keplerian velocity
$v_K(r)$.
    The gravitational acceleration
$\bf g$ diverges as
$r \to 0$, but of course the
presence of a star or black hole
changes this dependence.
    Instead of including the finite
size of the central object,
 the gravitational potential
is smoothed close to the origin,
$\Phi = - {GM}/(r^2+z^2+r_i^2)^{1/2}$,
where $r_i$ is the smoothing radius.
  The value $r_i$ is always much smaller than
the size of the computational region.
   For this smoothed potential,
the  Keplerian velocity (for $z=0$) becomes
$ v_K =r \sqrt{GM}/(r^2+r_i^2)^{3/4}$.
Our results do not depend significantly
on $r_i$ because the main part of the
outflow occurs from the inclined magnetic
field in the region of the disk where $r \gg r_i$.

\subsection{Numerical Method}

    Equations (24) - (26) and
(28) were solved with our Godunov
type numerical code (Koldoba et al. 1992;
Koldoba \& Ustyugova 1994;
Ustyugova et al. 1995).
    The code is
based on the ideas of Roe (1986)
for hydrodynamics and the related ideas of
Brio \& Wu (1988) for MHD.
  This type of TVD numerical scheme
has also been developed and investigated
by others
(for example, Ryu, Jones \& Frank 1995).
   The code has passed a
number of essential
tests (
Koldoba et al. 1992;
Koldoba \& Ustyugova 1994
)
which are analogous to those described  by
Ryu et al. (1995).
  Compared with our earlier applications of
this code  (Ustyugova et al. 1995;
Koldoba et al. 1996), a number of
improvements have been made,
including a procedure for
guaranteeing that
${\bf \nabla}\cdot{\bf B}=0$.
   To satisfy the condition
${\bf \nabla}\cdot{\bf B}=0$,
we projected the calculated
magnetic field to the sub--space of solenoidal
functions $\tilde{\bf B}$ at each time step.
  We introduced the function
$\Psi$, which satisfies the equation
$$
\left [ r \frac{\partial}{\partial r} \left ( \frac{1}{r}
\frac{\partial}{\partial r} \right ) +
\frac{\partial^2}{\partial z^2}
\right ] \Psi =
-~r \left ( \frac{\partial B_r}{\partial z} -
\frac{\partial B_z}{\partial r} \right )~.
$$
Then the magnetic field $\tilde{\bf B}$
was calculated for which
${\bf\nabla}\cdot\tilde{\bf B}=0$.
A similar method was used by
Ryu et al. (1995).

   Most of the simulations were done
on a grid with
$N_r \times N_z$ points in
cylindrical coordinates.
   For the
calculations on an approximately
square region we used
an inhomogeneous grid with $100 \times 100$
points while for the axially elongated
region we used a homogeneous grid
with $50 \times 200$ points.
 We also did a smaller number of simulations using
spherical coordinates $(R, \theta,\phi)$
and a grid $N_R \times N_\theta = 100 \times 50$.

\subsection{Initial Conditions}

   The motivation for this work was
the study of stationary MHD flows.
   Hence it may  appear that the initial conditions
are unimportant.
   However, in practice, an unfavorable choice
of initial conditions can lead to an essentially
longer stage of transition to stationary state,
or even worse, stationary flows may never be reached.
   The region right above the disk is the most
important, because the velocity distribution
near the disk determines the number of boundary
conditions (the flow may be subsonic or supersonic).
   Also, the physical parameters, such as density
and magnetic field, are largest just above the disk.
   Hence, we worked more carefully
on the equilibrium at small
values of $z$.
   At large $z$, approximate
equilibrium in the $z$-direction was sufficient,
because the magnetic field, which
is dominant in the corona, stabilizes matter
against the violent movements.
  The expressions given below
are found to be useful
initial conditions which
give a smooth start
up of the outflows.

   The initial conditions are arranged
as follows.
   The disk and corona are considered
to be threaded by a poloidal magnetic field
of monopole type (Sakurai 1987),
$ {\bf B}_p = { Q( {\bf R} - {\bf
R}_Q)}/{ |{\bf R} -{\bf R}_Q |^3}$,
where $Q=B_0 h^2$ is the
``charge'' of the monopole,
${\bf R}_Q$ is the position
vector of the monopole
located on the symmetry axis at a distance
$h$ below the disk.

   The temperature on the disk
surface, which is proportional
to the
square of (isothermal) sound
speed $c_T^2=p/\rho$,
was taken to have the dependence
$$
c_T^2 = \Phi(r,0)\left(\kappa +
\kappa_\ast e^{-r^2/{r_T^2}}\right)~,
\eqno(29)
$$
where $\kappa$ and $\kappa_\ast$
are parameters,
$r_T$  is a characteristic
radius inside of which the
 disk is relatively hot.
For specificity we take  $r_T=2 r_i$.
  For  $r\gg r_T$,
${c^2_T}/{\Phi(r,0)} \approx \kappa$;
that is, the sound speed is
constant fraction of the Keplerian
velocity in the region where we expect
centrifugally/magnetically driven
outflows from the disk.
  The term in (29) with $\kappa_\ast$
increases the temperature in the region
near the axis, where magnetic or centrifugal
outflows are not expected.
  For actual conditions,
this part of the flow may be
connected with the star or  black hole
(Livio 1997).

\begin{figure*}[t]
\epsscale{0.5}
\plotone{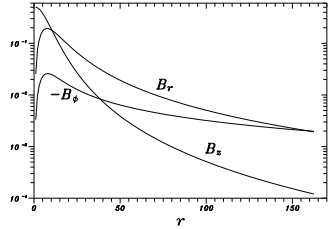}
\caption{
The figure shows the
radial dependences of the different
components of the  initial magnetic
field on the surface of the disk.
}
\label{Figure 4}
\end{figure*}

    We supposed that the initial
temperature of the corona is a
function only of $r$, so that the
equation (29) is the initial
condition for the entire
computational region.
   Also, we supposed that in $z-$direction
the gravitational force
is balanced by the pressure gradient,

$$
\frac{1}{\rho} \frac{\partial p}{\partial z} =
\frac{c_T^2(r)}{\rho}
\frac{\partial \rho}{\partial z} = - \frac{\partial
\Phi}{\partial z}~.
\eqno(30)
$$
The solution of this
equation is
$$
p(r,z) =p_d(r) \exp \left[ \frac{\Phi(r,0) -
 \Phi(r,z)}{c_T^2} \right]~,
\eqno(31)~,
$$
where $p_d(r)$ is the
pressure on the disk surface, and
 $\Phi(r,0)$ is the gravitational
potential on the disk.

    In the initial state, the surface of the
disk is in equilibrium.
   We suppose that the gravitational
force on the disk surface
is compensated by the centrifugal force,
while the matter pressure
gradient in $r-$direction is compensated
by the magnetic force.
  That is, on the disk ($z=0$),
$$
\frac{\partial p_d}{\partial r} + \frac{1}{8 \pi r^2}
\frac{\partial (r B_\phi)^2}{\partial r} = 0~.
\eqno(32)
$$
Solution of this equation for pressure on the disk
$p_d(r)$ and current
$I_d(r)=(c/2)r B_{\phi}|_{z=0}$ flowing through a
circular area of radius $r$
on the disk can be written as
$$
p_d(r) = \frac{B_0^2}{8 \pi} {\rm cos}^n\theta~,
\eqno(33)
$$
$$
I_d(r)={cB_0 h \over 2}
 \bigg[ \frac{n}{n-2} (1-{\rm cos}^{n-2}\theta)
-(1-{\rm cos}^n\theta)\bigg]^{1/2}~,
\eqno(34)
$$
where ${\cos}\theta=h/(h^2+r^2)^{1/2}$, with
$h=$const, and
$\theta$ is the inclination of magnetic field line
 to the axis of rotation.
This gave us possibility to find initial
azimuthal magnetic field along the disk,
$B_\phi(r)=2I_d(r)/cr$.  Close to the
disk, $rB_\phi$ is approximately constant
along a magnetic field line.
The three components of the
initial magnetic field on the disk are shown
in  Figure 4.

To escape rapid twisting of the magnetic
field
due to the difference
between the azimuthal velocities
of the disk and the corona,
we supposed that the corona
initially rotates with an angular
velocity which is constant on
cylinders $r=$const and equal to $v_K/r$ of
the disk.
As a result of this rotation the
corona is not in equilibrium in the
$r$-direction.
 However, this lack of initial equilibrium
does not disrupt the evolution of outflows
from the disk, and it does not affect
the final stationary states where
the flow reaches equilibrium.

\subsection{Boundary Conditions}

   The lower boundary of our
simulation region is the disk
which is perfectly conducting
and rotates at the Keperian rate.
   Thus the tangential component
of the electric field
in the system of coordinates
rotating with the disk
is zero,
$$[({\bf v} - {\bf v}_d)
\times {\bf B}]_{r,\phi} =0~,
\eqno(35)
$$
at $z=0$, where
${\bf v}_d = v_K {\bf e}_\phi$, and
$\bf v$ is the fluid velocity just
above the disk.
   This condition means that
in this system of coordinates
the poloidal velocity is parallel
to the poloidal magnetic field at $z=0$.

   The magnetic field is frozen into the disk
so that $B_z=B_{dz}(r)$,
where $B_{dz}$ is a given
function of $r$ and is determined by
the $z$-component of the initial monopole
magnetic field.
  Notice that the two other field components,
$B_r$ and $B_\phi$, are not fixed on the disk
and change with time so as to satisfy the
MHD equations in the computational region.

We suppose that
the density and entropy
 on the disk surface are fixed,
$\rho=\rho_d(r),~S=S_d(r)$ with $\rho_d(r)$
and $S_d(r)$ given functions of $r$
which follow from equations
(29) and (31).

  Note that in the present work, the
velocity of outflow from
the disk is a free variable.
This is different from our earlier
work (Ustyugova et al. 1995).
     When the velocity of outflow from the
disk is less than the slow
magnetosonic speed, then
the number of boundary conditions we
have is sufficient.
    However, if the outflow velocity is
super slow magnetosonic,
then there should be an
additional boundary condition.
   Because we do not have this additional
boundary condition, we
suppose that the amplitudes of
the correponding
outgoing waves are equal to zero.
   This is equivalent
to the fact that we use the values of
calculated parameters in the cells just
above the disk.

  On the $z-$axis, all fluxes
normal to this axis are equal to zero.
  On the outer boundaries,
$r=R_{max}$ or $z=Z_{max}$,
the ``free'' boundary
conditions $\partial F_j/{\partial n}=0$
were used
for all variables excluding $B_\phi$.
Here, $\partial/{\partial n}$ is  the derivative
perpendicular to the boundary,
$F_j=\{\rho,~ f(S),~ v_r,$
$v_\phi,~ v_z,~ B_r,~ B_z\}$.
  Our earlier simulation study
(Romanova et al. 1997) showed that
the condition
$\partial B_\phi/{\partial n}=0$
can lead to unphysical results.
The outer boundary condition on
$B_\phi$ is considered in detail in \S 4.

\section{Influence of Boundary Conditions
on Flow}

    If the process of outflow formation
is strongly non-stationary,
then the problem of the influence
of outer boundary
conditions may not appear.
    This is because it is
difficult to separate
the influence of boundary conditions
from effects connected with non-stationarity.
  However, when the flow
goes to a steady-state,
we observed that the stationary flow pattern
can depend on the imposed  outer
boundary conditions
and in some cases on the
shape
of the simulation region.

   It is important to
eliminate the influence of boundary conditions.
   It is possible, if (1) the flow is supersonic
(super fast magnetosonic)
and it is {\it  perpendicular} to the outer
 boundaries (then information flows out
 of the simulation  region), or
(2) the correct boundary
conditions are chosen by some method.
  The first condition cannot be realized
during the stage of
 establishing of the flow, because
initially, the flow is subsonic.
   If the flow is supersonic,
but is not perpendicular
to the boundary, then the Mach cones
may be partially directed inside the
simulation region and even supersonic flow
may influence the flow inside the
region.
  The orientation of the Mach cones
depends in general on the
shape of simulation region.

  The second condition can be realized only in
some approximation.
  The ``best'' outer boundary conditions
are those which influence  only the
vicinity of the boundaries and not
the central part of the simulation region.
  This  involves  all flow variables,
but we will discuss only the outer
boundary condition on $B_\phi$, because
we found that this condition
had the strongest influence on the
calculated flows.

The final flow may depend
on both the Mach cone
orientation at the
boundaries (shape of the region)
and on the outer boundary
condition on $B_\phi$.
   In different situations
one of these factors may be more
important than the other.
    To separate  their
influence on the final flow
pattern, we discuss in
\S 4.1 simulations for a fixed
simulation region, but with different
outer boundary conditions on $B_\phi$.
   Next, in \S 4.2 we fixed
the boundary condition
on $B_\phi$,
and investigated the
dependence of the flow on
the shape of simulation
region and the
Mach cones orientation
at the outer boundaries.
  In \S 4.3 we discuss both factors.

\subsection{Dependence of Flows
on Outer Boundary Condition on $B_\phi$}

Here, we present results of simulations
for a fixed elongated simulation region
$R_{max}=50 r_i$, $Z_{max}=200 r_i$
for three different outer boundary conditions
on $B_\phi$: (1) a standard ``free''
boundary condition, (2) a``force-free''
 boundary condition, and
(3) a ``force-balance''
boundary condition.

\subsubsection{``Free'' Boundary
Condition}

   First, we performed simulations for the simplest
standard
``free'' boundary condition on $B_\phi$,
$\partial B_\phi/\partial n=0$.
  We observed that this boundary condition
may give an artificial
force on the boundary
which influences the flow
within the computational region.
  For example, if we suppose that on the top boundary
$\partial B_\phi/\partial z=0$, then the
radial component of the current-density equals to
zero,
$j_r=
-(c/4\pi){\partial B_\phi}/{\partial z}$ $=0$,
which means that the poloidal current-density has only
a $z$-component ${\bf j}_p=j_z\hat{\bf z}$.
  This means that the poloidal
current-density ${\bf j}_p$ is not parallel
to the poloidal magnetic field ${\bf B}_p$.
   Consequently,  there is
a force (density) ${\bf j}_p\times{\bf B}_p/c\neq 0$
acting in the $\phi$ direction,
opposite to the rotation of the disk.
Figure 5 shows the geometry.

%\placefigure{fig5}

\begin{figure*}[t]
\epsscale{0.4}
\plotone{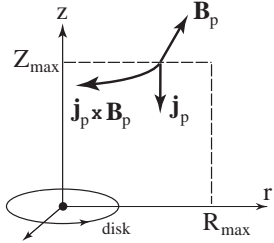}
\caption{
The figure demonstrates the artificial
force which can appear on the top outer
 boundary of the simulation region in the case
of a ``free'' boundary condition
on $B_\phi$.
  Here, ${\bf B}_p$
and ${\bf j}_p$ are the poloidal magnetic field and
current-density (filled vectors).
The hollow arrow shows the artificial force
acting in the azimuthal direction.
}
\label{Figure 5}
\end{figure*}

  These `boundary' forces
act such way that the flow
never reaches a stationary state.
  To check this fact, and to be sure that
this is not an effect of
non-stationarity of our initial
configuration, we did
simulations for cases which
went to a stationary
state with other outer
boundary conditions.
  After establishing stationarity,
we substituted the outer boundary
conditions on $B_\phi$ to a
``free'' boundary condition.
   We observed that the stationary state was
destroyed for the  reasons mentioned above.
   Figures 6a,b
demonstrate one stage of this destruction,
when the poloidal velocity decreased and became
less than fast magnetosonic speed in all of
the computational region.
  Even the fluxes of mass
and other physical parameters
through the boundaries are not constants in
 this simulation.
  Also, matter with magnetic
flux enters the region
from the right-hand side, which is due
to the flow being  sub-fast magnetosonic.

  To avoid this artificial force, we proposed
a ``force-free'' outer boundary condition on
$B_\phi$ (Romanova et al. 1997)  which
we discuss in the next subsection.

\subsubsection{``Force-Free'' Boundary
 Condition}

Another possibility to
consider is that the toroidal
component of the
magnetic force is
zero on the outer boundaries.
That is,
${\bf j}_p\parallel{\bf B}_p = 0$
on the outer boundaries.
  We can write this condition as
$$
   {\bf B}_p\cdot{\bf \nabla}( r B_\phi)=0~.
\eqno(36)
$$
  We performed simulations
with this boundary condition
in the elongated region and observed that
the flow reached a
stationary state (see Figures 6 c,d).
 This flow has many characteristics of stationary
flow. Fluxes of mass, energy, and momentum,
integrated over different cross-sections,
are constants.
   Integrals of motion
along magnetic field lines
 are also constants.
   The flow is {\it well collimated} inside the
simulation region (see Figures 6c, d).
However, more detailed analysis (see \S4.2)
shows that this collimation is
artificial.
  The ``force-free'' boundary condition for $B_\phi$
is superior to the ``free''
boundary condition, because it leads to
a stationary state, but it
does not give the physically correct flow.
   In reality, the magnetic
force should not be zero on the boundary.
  There is a magnetic  force
pushing matter outward through
the outer boundaries.
 One can see from Figure 6d

that the poloidal current-density
(dashed lines) is not parallel
 to the poloidal magnetic field (solid lines).
    However, on the boundaries (Figure 6d)
the two vectors are forced to be parallel
and thus the poloidal force equals  zero.
   This boundary condition is better than
the ``free'' boundary condition in
the sense that there is no strong artificial
force at the boundary.
   From the other side, when we
put the force equal to zero, it is analogous
to application of a force  equal to
the real force but with the opposite sign.

   This is one of the factors which may
lead to artificial collimation.
Another possible factor (Mach cones orientation)
depends on the shape of the simulation region
and is discussed in \S 4.2.

\subsubsection{``Force-Balance'' Boundary
Condition}

  As a next step for improving
the outer boundary
condition on $B_\phi$, we
take into account the fact that the
magnetic field is not
force-free and ${\bf j}_p$ is
not parallel to ${\bf B}_p$.
   We start from equation (16) for $B_\phi$
and write it in the form
$$
     r B_\phi=\Omega \sqrt{4\pi \rho_A}~
  {{r^2-r_A^2}\over{1-{\rho_A/\rho}}}\approx
 - \Omega ~\sqrt{{4\pi}\over \rho_A} ~{\rho r^2}~,
\eqno(37)
$$
where we assume that the density at the boundary
is much less than that at the Alfv\'en surface,
$\rho \ll \rho_A$ for $r^2 \gg r_A^2$.
  Then, we obtain
$$
{\bf B}_p\cdot {\bf \nabla} (r B_\phi) =
 -\Omega ~\sqrt{4 \pi\over \rho_A}~
{\bf B}_p\cdot{\bf \nabla} (\rho r^2)
$$
$$
=r B_\phi ({\bf B}_p \cdot {\bf \nabla}) { \ln}
 (\rho r^2)=\alpha B_r B_\phi~,
\eqno(38)
$$
where we supposed that $\rho r^2 = F(\Psi) r^\alpha$
and took into account that
$\Omega$ and $\rho_A$ are constants
along magnetic field lines.

\begin{figure*}[t]
\epsscale{1.0}
\plotone{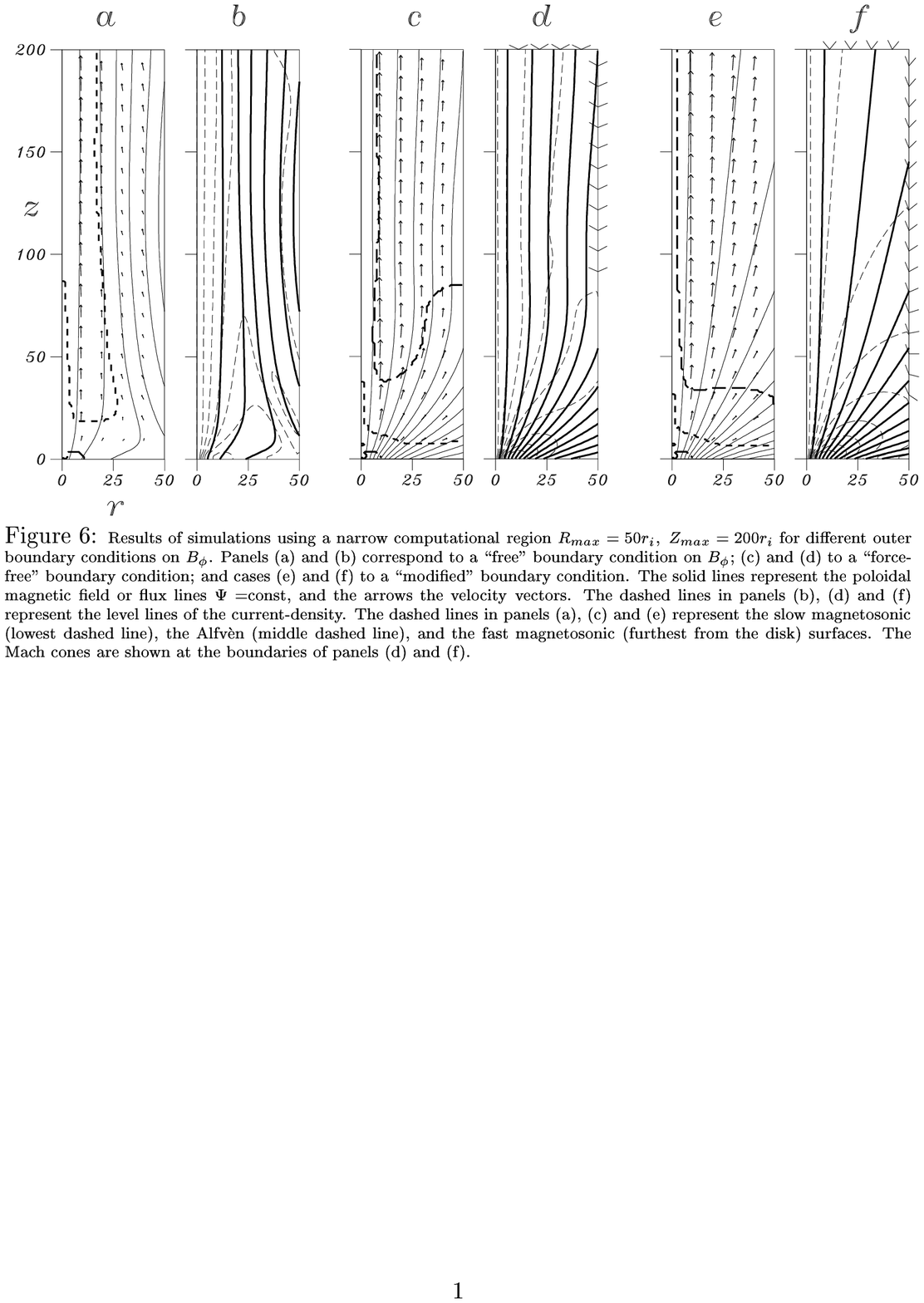}
%\caption{ Results of simulations using
%a narrow computational
%region $R_{max}=50 r_i,~Z_{max}=200 r_i$
%for different outer boundary conditions on $B_\phi$.
%Panels (a) and (b) correspond
%to a ``free'' boundary condition on $B_\phi$;
% (c) and (d) to a ``force-free''
%boundary condition; and
%cases (e) and (f)  to a
%``modified'' boundary condition.
%  The solid lines represent the
%poloidal magnetic field
%or flux lines
%$\Psi=$const, and the arrows the velocity vectors.
%The dashed lines in panels (b), (d) and (f) represent
%the level lines of the current-density.
%   The dashed
%lines in panels (a), (c) and (e) represent the
%slow magnetosonic (lowest dashed line), the
%Alfv\`en (middle dashed line),
%and the fast magnetosonic
%(furthest from the disk) surfaces. The Mach cones
%are shown at the boundaries of panels  (d) and (f).}
\label{Figure 6}
\end{figure*}

Finally, we obtain the outer
 boundary condition as
$$
    {\bf B}_p \cdot {\bf \nabla} (r B_\phi)
=\alpha B_r B_\phi,
\eqno(39)
$$
where $\alpha$ is a parameter.
  In this case we got stationary
flows which are {\it not collimated}
 in the simulation region (see Figures 6e, f).
Fluxes through the outer surfaces
and integrals along magnetic field
lines are well conserved, as in the
 case  of collimated flow, described
in \S 4.1.2.

%\newpage

   The question arises, which boundary condition is
correct, ``force-free'' or ``force-balance'' ?
  The ``force-balance'' condition is clearly the
physical condition because it does not
generate an artificial force on the boundary.
   However, it is more difficult to apply
because there is no direct method
for determining the parameter
$\alpha$.
  It can only be obtained iteratively
using additional simulations, which
is very time consuming.
   Our analysis indicates that the ``force-free''
boundary condition gives good results
as compared with those obtained
using the ``force-balance'' condition {\it if}
the shape of the simulation region is
not elongated in the $z-$direction.

Below, we investigate
different runs for
``force-free'' outer boundary
conditions on $B_\phi$, but
for different shapes of the simulation region.

\subsection{Dependence of  Flows
on Shape of  Region:  Orientation of Mach Cones}

   We noticed empirically that
results of simulations
depend significantly on the
shape of simulation region.
   The ratio between $R_{max}$
to $Z_{max}$ is critical.
    We observed that when the region is
elongated in $z-$direction,
then the flow has tendency to
collimate.
  When the region is square, or spherical, or
elongated in $r-$direction, then the outflow is almost
spherical, that is,
only slightly collimated.
   Here, we present results of simulations all
with ``force-free'' outer boundary conditions but
different shapes of the simulation region.

   First, we investigated
the case where the height
of the region is the same
as before, $Z_{max}=200 r_i$, but
the region is much wider, $R_{max}=170 r_i$.
   Figure 7   shows that
in this case we got almost spherical outflow,
which is  very
different from the well-collimated
outflow in the narrow region
at the same boundary conditions
(Figures 6c, d).

\setcounter{figure}{6}
\begin{figure*}[t]
\epsscale{1.0}
\plotone{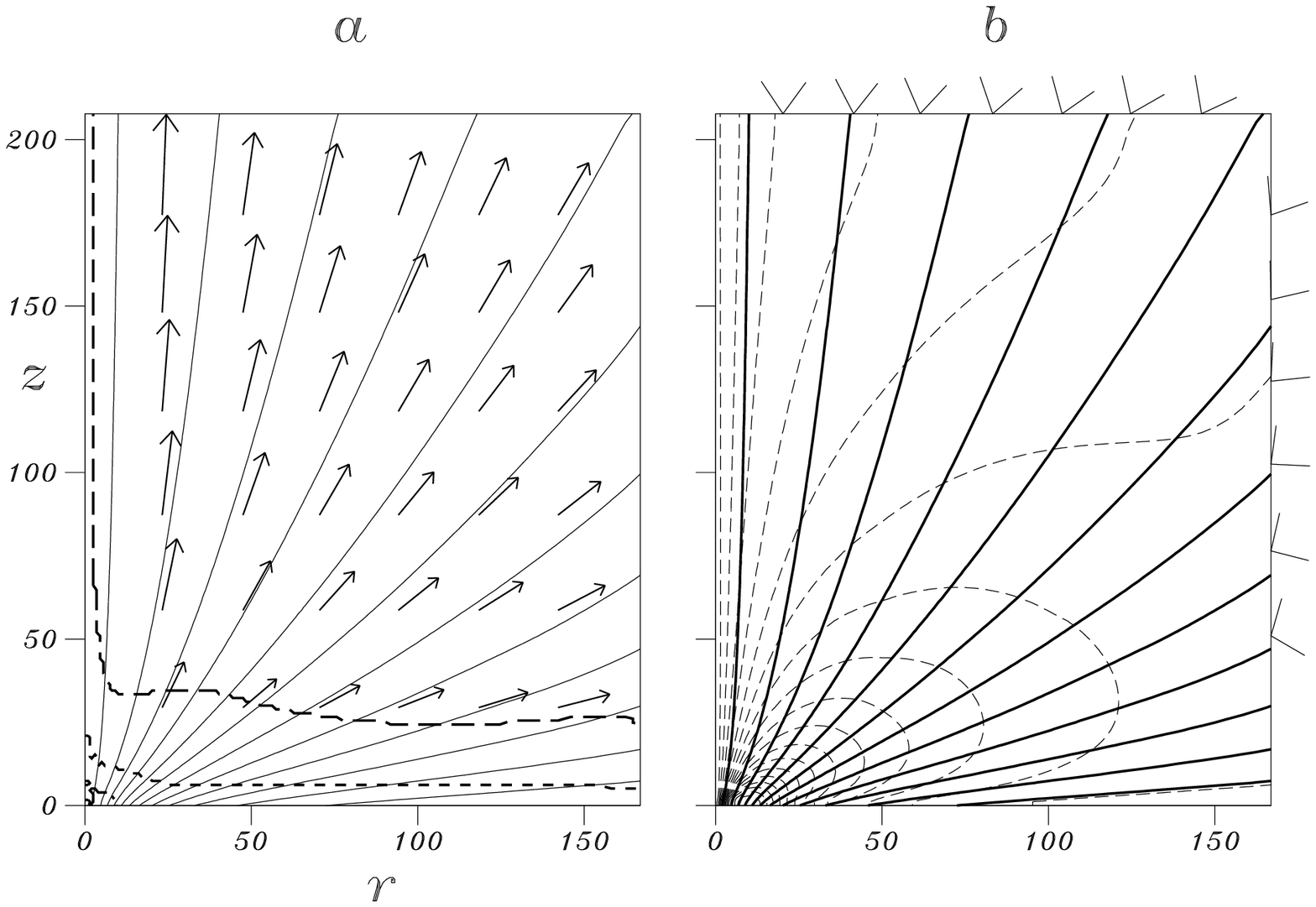}
\caption{Results of simulations
for an approximately square
simulation region
$R_{max}=170 r_i,~Z_{max}=200r_i$
using the
``force-free'' outer boundary condition on $B_\phi$.
Here, $r$
 and $z$ are measured in units of $r_i$,
which is the inner radius of the disk.
The lines and arrows have the same meaning
as in Figure 6.}
\label{Figure 7}
\end{figure*}

   We also performed similar
simulations in spherical
coordinates with $R_{max}=170 r_i$,
and got similar result.
   The question is why the flows
are so different for different shapes of
the simulation region?
   In all cases the flow is
super fast magnetosonic
in most of the region.
    However,  note
that even if the flow is super fast magnetosonic,
information can flow in from the boundaries to
the simulation region, if the Mach cones
are directed inside the simulation region.

%\placefigure{fig7}

 The Mach cone projected
onto the poloidal plane has
a half opening angle
$\varphi$ which is
 $$
 \tan^2\varphi
=\frac{(v_A^2 + c_s^2)(v^2_p - v_c^2)}
{(v_p^2-c_{sm}^2)(v_p^2-c_{fm}^2)}~,
\eqno(40)
$$
where
$c_{sm}$ and $c_{fm}$  are the
 slow and fast magnetosonic velocities, respectively,
which satisfy $c_{s,fm}^4-c_{s,fm}^2(c_s^2+v_A^2)
+c_s^2 v_{Ap}^2=0$ (with $v_A^2 ={\bf B}^2/(4\pi\rho)$
and $v_{Ap}^2={\bf B}_p^2/(4\pi\rho)$)
and
$v_c \equiv  v_{Ap} c_s/(v_A^2 +c_s^2)^{1/2}$
 is the ``cusp'' velocity (Polovin \& Demutskii 1980;
Lovelace et al. 1986;
Bogovalov 1997).
   Figures 6 - 8 show the Mach cones
on the outer boundaries for
different shapes of the simulation region.
   We find that in the case of an elongated
region  (Figure 6d) an essential part of  the Mach
cones is directed into the simulation region,
whereas in the case of an
almost square region (Figure 7b)
only very small part  of the Mach cones is
directed into the  region.
  Figure 8 shows that the  most
desirable geometry - where
information flows outward across
the outer boundary - is obtained in
spherical coordinates
where all Mach cones are directed outward
from the simulation region.

\begin{figure*}[t]
\epsscale{1.0}
\plotone{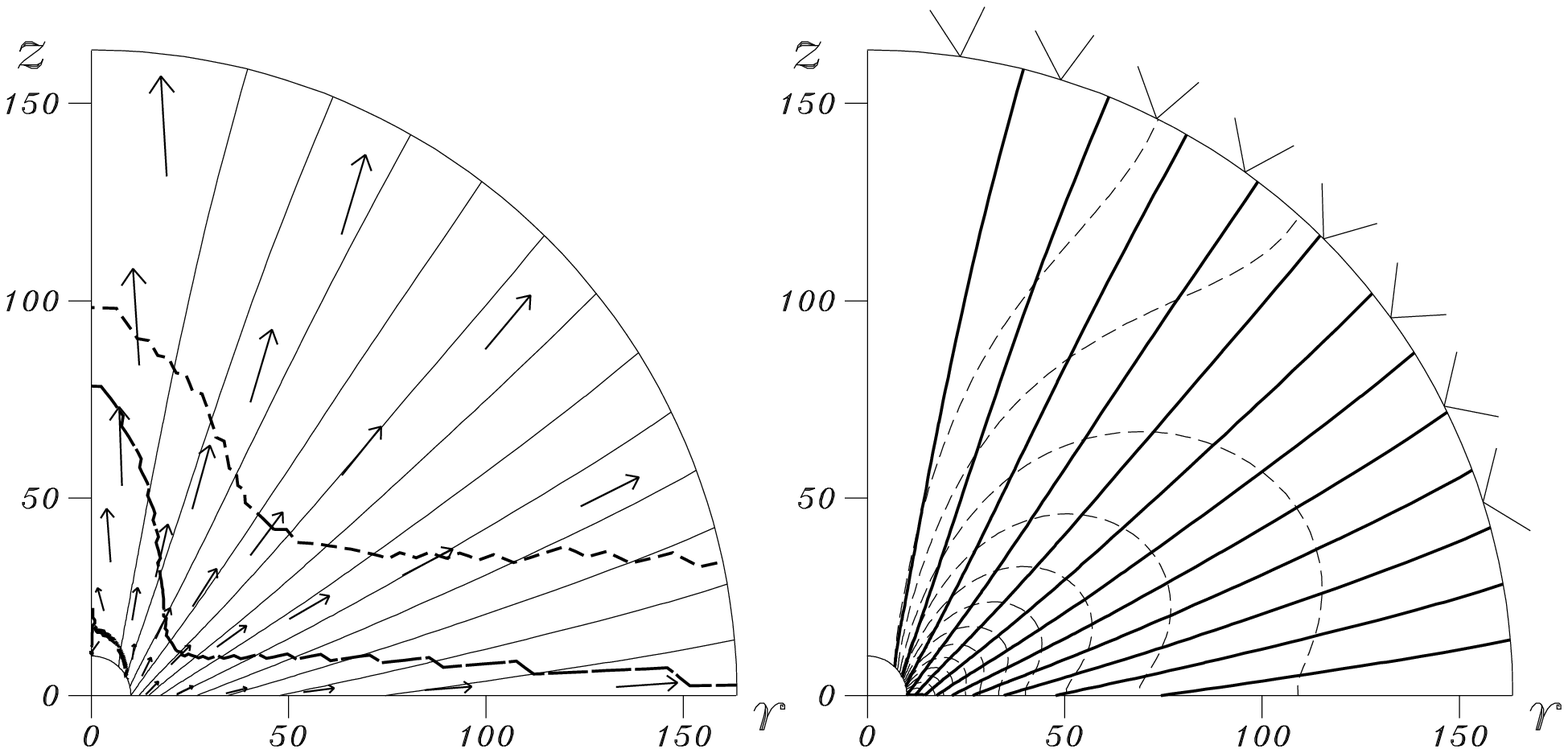}
\caption{Results of simulations
for the same case as Figure 7 but
using a spherical coordinate system
and simulation region.}
\label{Figure 8}
\end{figure*}

   The elongated region is the least
desirable and as discussed it gives
artificial collimation
of the flow.
  Note that the first
stationary MHD flow solutions
(Romanova et al. 1997) were obtained
using a simulation region elongated
in the $r$-direction.

\subsection{Discussion of Boundaries}

   Regarding the outer boundaries,
we conclude that simulated flows may
depend on both the
outer  boundary condition on $B_\phi$
and on the shape of simulation region
 (the Mach cone orientation on the
outer boundary).
   The influence of each of these
factors may be different
in different situations.

   The orientation of the Mach cones
at the boundary is not connected directly
with existence and configuration of a stationary
flow.
  However, if   Mach cones are partly directed
inside the simulation region along
part of the outer boundary, then the question
arises: what is the result of this influence,
and how strong is it?
  Our simulations  shown that
for a ``force-free'' boundary condition
on $B_{\phi}$ the result
is artificial collimation of the flow,
whereas in the case of a ``free'' boundary condition
there is destruction of a stationary flow which
was arranged as an initial condition.

   From comparison of cases
shown in Figures 6d and 7b
it is not clear that Mach
cones are responsible for the
collimation of the flow in the
case shown in Figure 6d.
   In the narrow region,
the magnetic field
at the right-hand, outer
boundary is much stronger
than in the case of wide region,
so that influence
of the non-exact ``force-free''
boundary condition
should be stronger in the
case of a narrow region.
   To check this possibility,
we performed simulations
in a small square region
with $R_{max}=50 r_i$ and $Z_{max}=50 r_i$
and found uncollimated
almost spherical outflow.
  This indicates that
the shape of the region
is the most important factor
affecting collimation.

Another question are evident.
   Why in the case of the
``force-balance'' boundary condition
on $B_\phi$ in the elongated region do
we find the physically correct
stationary solution?
   We suggest that during establishment
of the stationary flow, which may be quite
violent (in spite of almost
stationary initial conditions),
this boundary conditions kept approach
to stationarity less violent (than in the case of
``force-free''
 boundary conditions) and kept the Mach cones
directed outward most of the time.
   The fact that in this case a
small part of the Mach cones is directed
inside the region, means that some
inflow of information
from the outer boundary may
have only a small affect on the flow.

 For some purposes, such as the study
of the propagation of
jets, it is attractive to use
a long  narrow computational region.
   The general conclusion of this section
is that a narrow region can lead to
artificial collimation of the flow
or invalid solutions unless
 special care is given to the
boundary condition on $B_\phi$.

\section{Stationary Flows:
Comparisons of Simulations with Theory}

    Here, we describe results of
simulations for a region $R_{max}=170 r_i$
and
 $Z_{max}=200 r_i$ with a
``force-free'' outer
boundary condition.
   Simulations of the flow in the same region but
with the  ``modified'' boundary condition
gave similar results.

    Figure 9 shows the
initial distribution on the disk of
the Keplerian velocity $v_K$, the
poloidal Alfv\'en velocity $v_{Ap}$,
and the sound speed $c_s$.
   Matter outflowing from the disk has a
time-independent
distribution of density as a function
of radius.
  The velocity
of outflow from the disk is
determined by the solution of the
MHD equations in the simulation region.
  The simulations  show that in a stationary state
matter just above the disk has a velocity
somewhat larger than the slow magnetosonic velocity.
This is in accord with the theory which
indicates  that the
slow magnetosonic surface is located inside
the disk (Lovelace, Romanova, \& Contopoulos 1993).

\begin{figure*}[t]
\epsscale{0.6}
\plotone{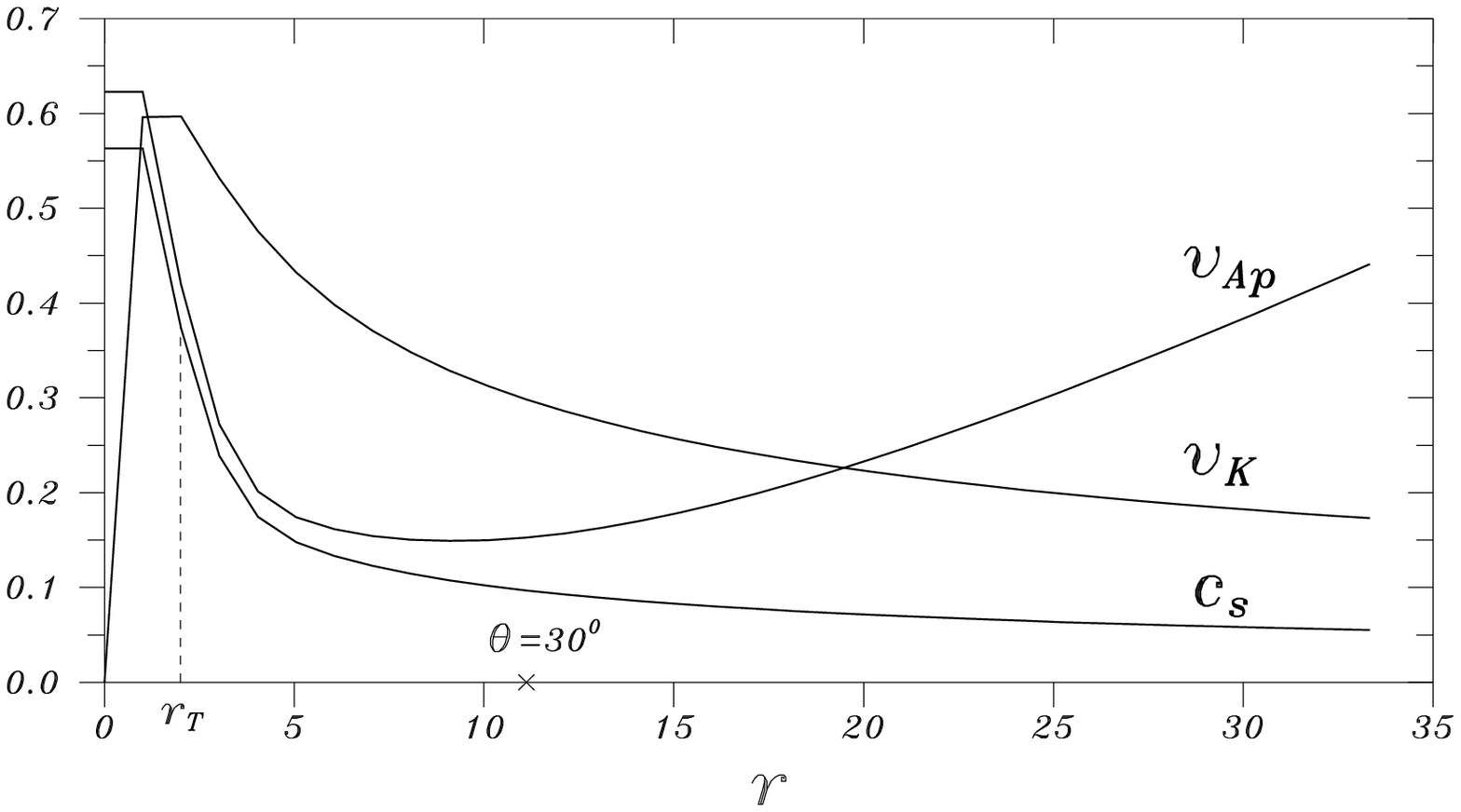}
\caption{Dependencies of velocities
on $r$ (in units of $r_i$)  along the disk in
the stationary state, where $r_i$ is
the inner radius of the disk.
 Here, $v_K$ is the Keplerian
velocity, $c_s$  the
 sound speed, and
$v_{Ap}$ the poloidal Alfv\'en velocity,
measured in units of $v_i$.
  The point $\theta=30^o$ shows
the location on the disk where
the poloidal magnetic field
is inclined to the $z-$axis
at an angle  $\theta=30^o$;
for larger $r$ we have $\theta>30^o$.
   We show only  $r \leq 35r_i$, because
the magnetic field lines
which start at larger $r$ are
highly inclined away from the $z-$axis
and also do not pass through
the fast magnetosonic surface within
our simulation region (see Figure 7).
   Inside the radius $r_T$
the disk is relatively hot
(see equation 29).}
\label{Figure 9}
\end{figure*}

  Moving away from the
disk, matter starts from a low velocity, is
gradually accelerated and crosses the
Alfv\'en and fast magnetosonic surfaces
(see Figure 7a).
  These surfaces are almost
parallel to the disk.
   Figure 10a shows the
variation of different velocities
along a representative magnetic field line,
  the third line away from the
$z-$axis in Figure 7, on the flow distance
$s$ from the disk.
   This field line, which
we refer to as the ``reference''
field line, crosses the top boundary
at about the midpoint of this boundary.
   This line is not special, but
it is inclined sufficiently to
the axis  that magnetic/centrifugal
forces are important.
   Figure 10a  shows in particular the
dependence of the poloidal velocity
$v_p(s)$, which becomes
larger than the Alfv\'en velocity $v_{Ap}$
fairly close to the disk,
at $s > 10 r_i$.
   Further, $v_p$ becomes
larger than the local escape velocity
$v_{esc}$ for $s> 25 r_i$.
   At larger distances, $v_p$ becomes larger
than the fast magnetosonic velocity
$c_{fm}$ at $s > 40 r_i$, and it approaches
an asymptotic speed which is about $1.75$
times $c_{fm}$ at the outer boundary of the
simulation region.
  The poloidal velocity is
parallel to the poloidal magnetic field
to a good approximation
in accord with the theory.
  Figure 10b shows the
dependences of $v_p(s)$ for different
field lines.

   Within the simulation region, the outflow
accelerates from thermal velocity
to a much larger asymptotic poloidal flow
velocity of the order of $0.5\sqrt{GM/r_i}$.
  Thus, the {\it acceleration distance}
for the outflow, over
which the flow accelerates from $\sim 0$
to, say, $90\%$ of the asymptotic speed, occurs
at a flow distance of about $80 r_i$.

\begin{figure*}[t]
\epsscale{0.5}
\plotone{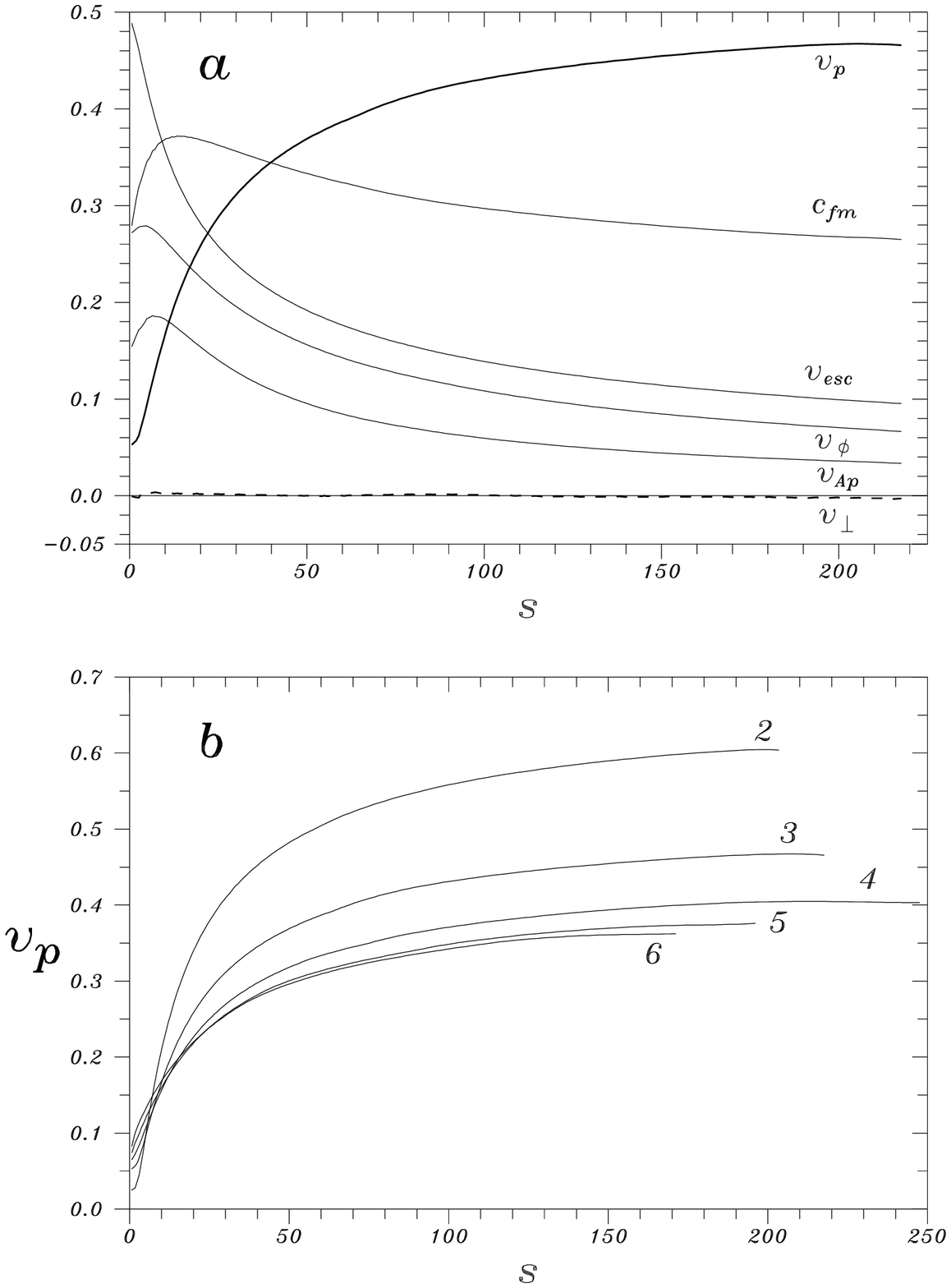}
\caption{
The top panel
($\bf a$) shows the dependences of different
velocities on distance $s$ measured
in units of $r_i$
from the disk along  the ``reference''
 magnetic field line.
   The velocities are measured in
units of $v_i$.
   The ``reference'' field
line is the third  field line
away from the $z-$axis in Figure 7a.
    This field line
crosses the top
boundary about in the middle.
  This field line ``starts'' from
the disk  at $r \approx 6r_i$
where it has an angle
 $\theta \approx 28^o$ relative to
the $z-$axis.
   Here, $v_p$ is the poloidal velocity along
the field line and
$v_\perp$ is the poloidal velocity perpendicular
to the field line.
Also, $v_{Ap}$  is the
poloidal Alfv\'en velocity,  $c_{fm}$
is the fast magnetosonic velocity, and
$v_{esc}$ is the local escape velocity.
   The bottom panel (${\bf b}$) shows the
dependences of $v_p(s)$ for different
field lines indentified by the number
of the field line counted
away from the $z$ -- axis in Figure 7a.}
\label{Figure 10}
\end{figure*}

\subsection{Mechanism of Acceleration}

   Figure 11 shows the different
forces acting   along
the ``reference'' magnetic field line.
      The centrifugal force ($F_C$) is larger
 than the magnetic ($F_M$)
 or pressure gradient force ($F_P$)
immediately above the disk
 $s < 10 r_i$.
  The magnetic force is few times larger
than the centrifugal force
 for larger distances, $s>10r_i$.
 Note that the pressure gradient
force is negligibly small.
  Thus, the main driving forces
pushing matter outward are
magnetic and centrifugal.

%\placefigure{fig11}

\begin{figure*}[t]
\epsscale{0.5}
\plotone{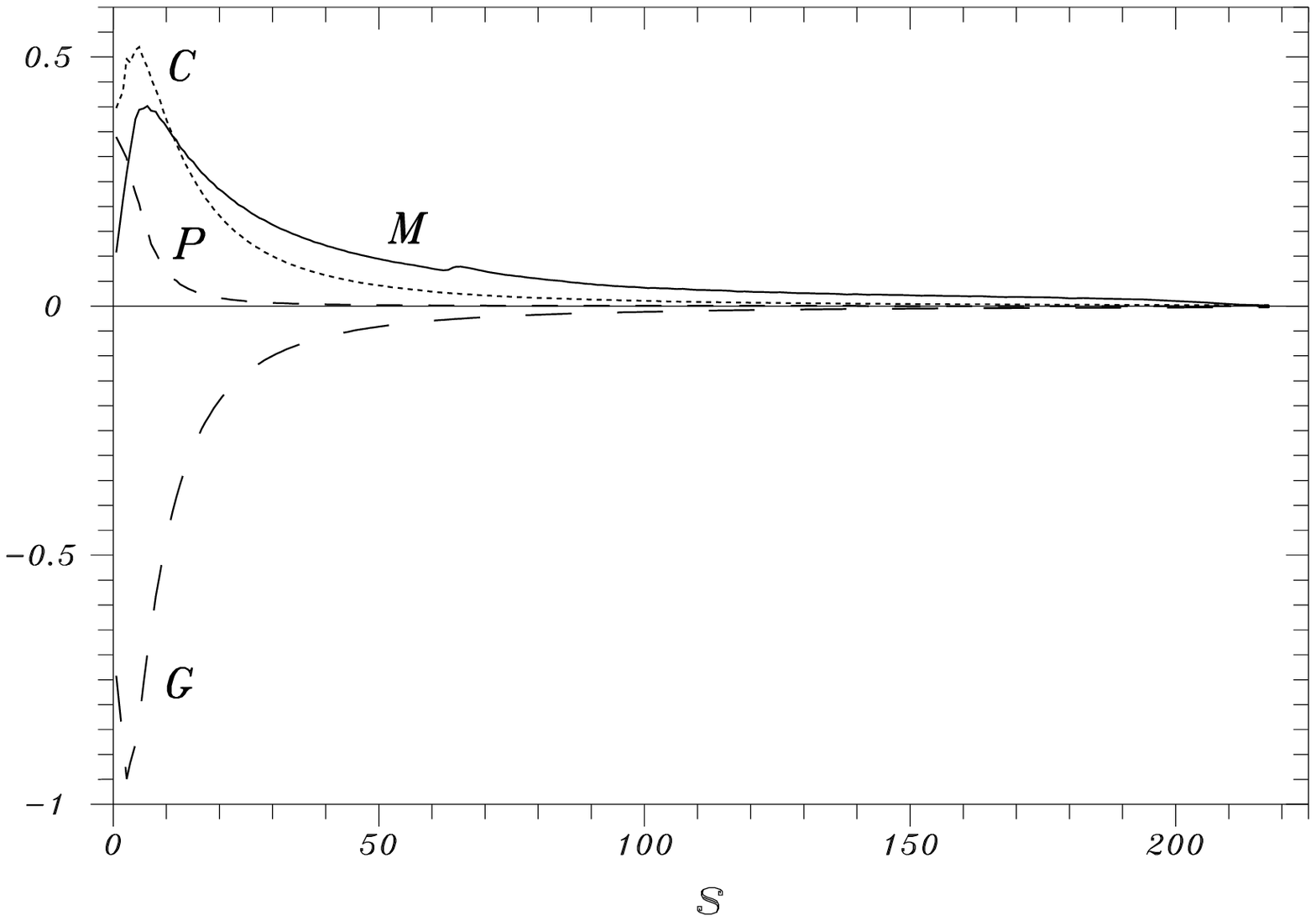}
\caption{
Forces acting along the
``reference'' field line.
 $M$ is the magnetic force,
$C$ the centrifugal force,
$P$ the pressure gradient force, and
$G$ the gravitational force.
The distance $s$ is measured in units of $r_i$.
The scale for forces is arbitrary.
}
\label{Figure 11}
\end{figure*}

    Each poloidal magnetic
field line is labeled by
its $\Psi$ value, which equals the magnetic flux
through the circular
region between the axis and the field line.
$\Psi$ increases from  zero on the axis
to a largest value on the field line
most distant from the axis.
  We integrated the forces to
obtain the total work performed
by the magnetic, centrifugal and other forces,
 along different field lines from the disk
to the outer boundary.
    Figure 12 shows the
dependence of this work on $\Psi$.
One can see that near the axis (small $\Psi$)
the main work is performed by the centrifugal force,
while the magnetic force is also important.
The work along the ``reference'' field line
marked by  ``$R$'' on the $\Psi$ axis,
is done mainly by the magnetic force
with the  centrifugal force  also important.
Going away from the axis to larger $\Psi$
and more
inclined magnetic field lines, one can see
that the magnetic force is
more and more important role.
Note that the work done by the
pressure gradient is small for all field
lines.

%\placefigure{fig12}

\begin{figure*}[t]
\epsscale{0.5}
\plotone{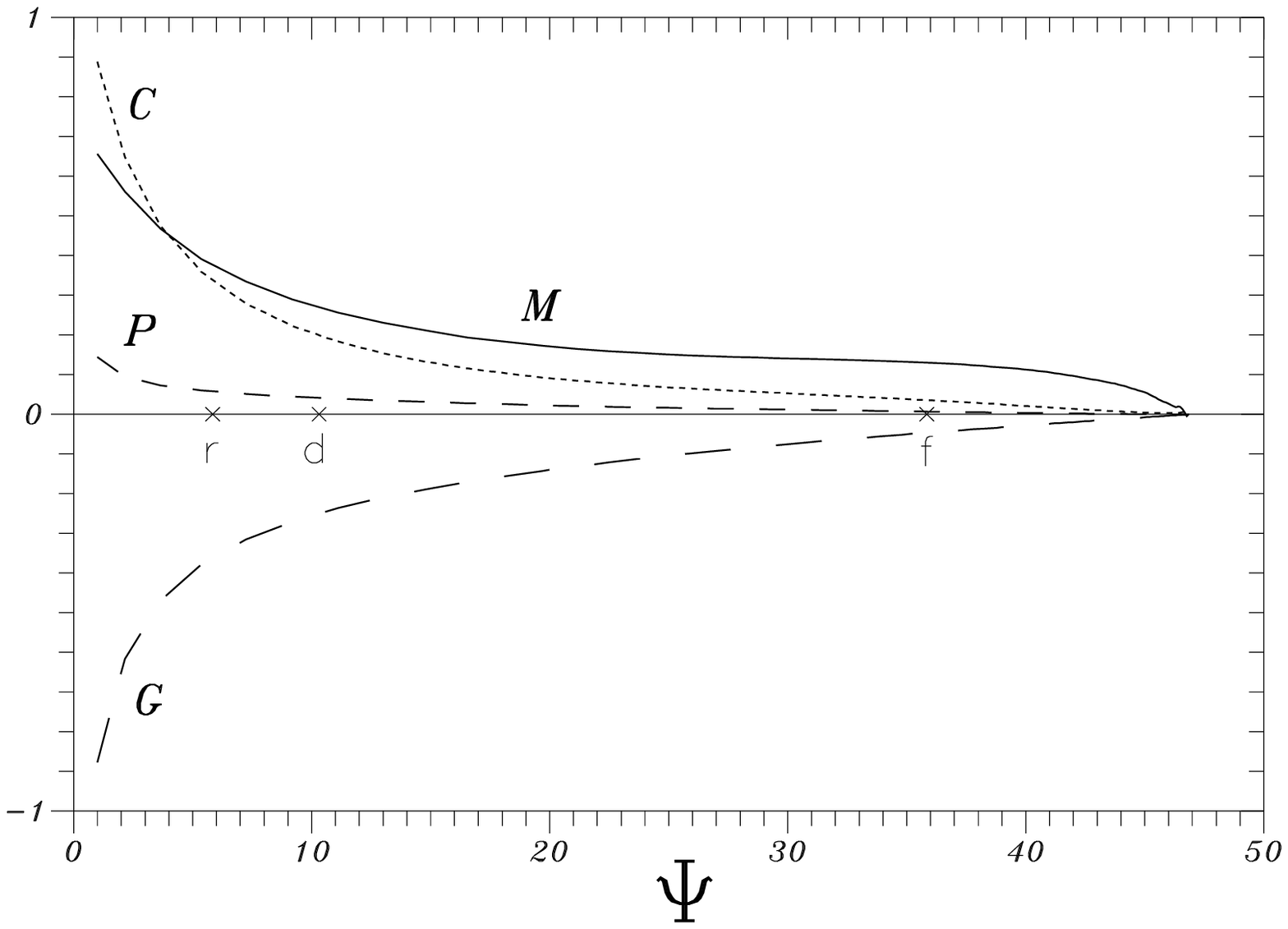}
\caption{
Work done by the different forces
along different field lines from the
disk to the outer boundary.
  Each field line
is labeled by its value of $\Psi$.
  The field line corresponding
to our ``reference'' line
is marked ``$r$,''  and the ``diagonal'' field
line which goes
through the top right corner
of the simulation region  is
marked ``$d$.''  The point ``$f$'' is
the largest radius at which the
flow goes through the fast
magnetosonic surface.   The letters $M$,
$C$, $P$, and $G$ stand for magnetic,
centrifugal, pressure gradient, and gravity.
}
\label{Figure 12}
\end{figure*}

\subsection{Analysis of Stationarity}

  A first indication of stationarity is
when the fluxes of mass
and other physical quantities
 become constants in time.
        We observed that the
 fluxes of mass ${\cal F}_M$ , angular
 momentum ${\cal F}_L$, and energy
 ${\cal F}_E$ calculated through the
middle of the region $z=0.5 Z_{max}$
become constants after about $t > 200 t_i$, where
 $t_i \equiv 2 \pi r_i/v_i$ where $v_i\equiv
\sqrt{GM/r_i}$ and $r_i$
is the inner radius of the disk.
  The time dependence of the
fluxes is shown in Figure 13.
    We performed simulations for much longer
 times, $t\sim 3500 t_i$,
and observed that these fluxes
were accurately  constants in time.
   Note that this time corresponds
to only $t=1.6 t_{out}$, where
$t_{out}=t_i \left( {r_{out}}/{r_i}
\right)^{3/2}=2216 t_i$.
    This indication of stationarity
is necessary, but not a sufficient
 sign of a valid stationary
MHD flow.

%\placefigure{fig13}

\begin{figure*}[t]
\epsscale{0.5}
\plotone{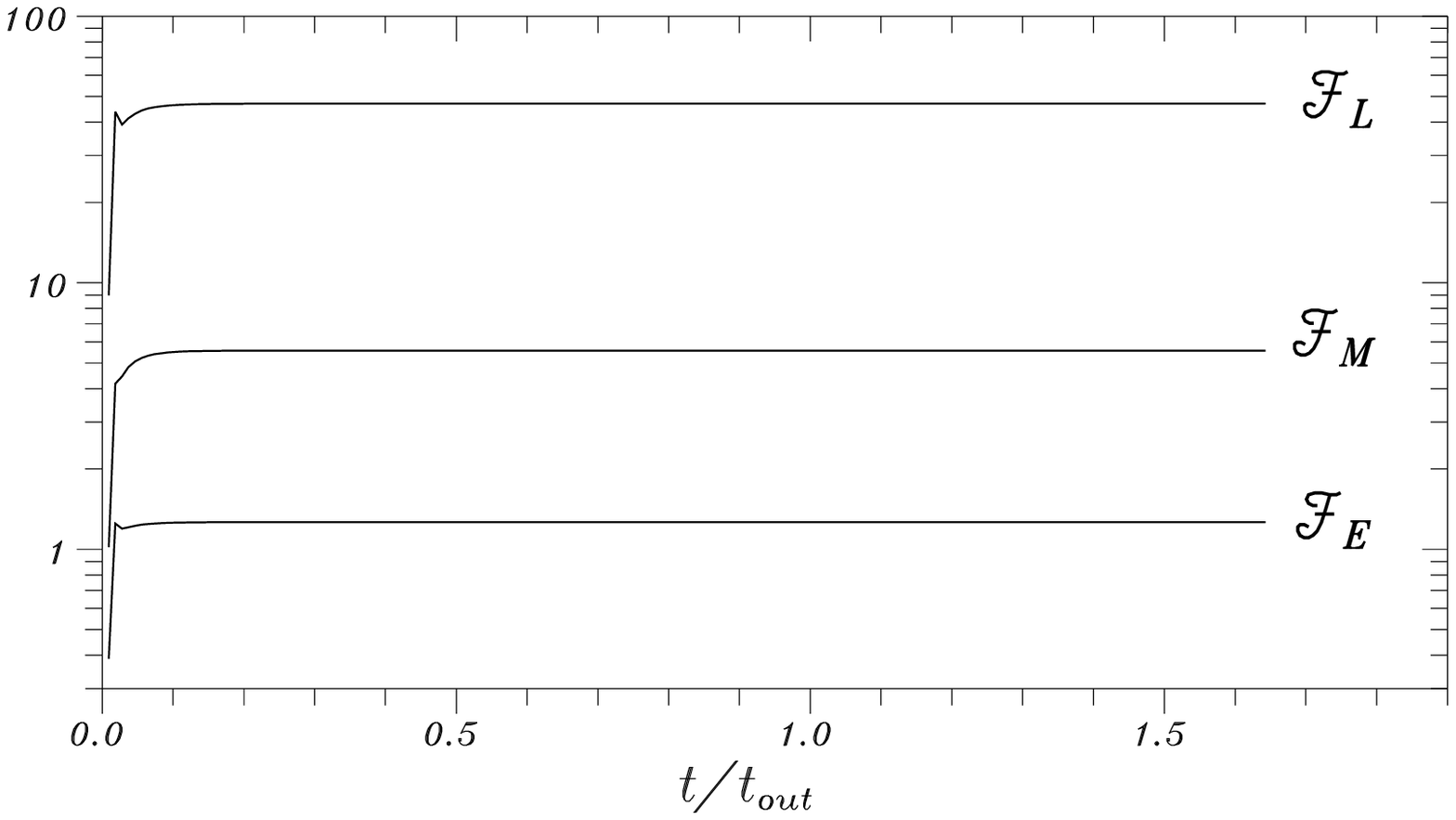}
\caption{Fluxes of mass ${\cal F}_M$,
angular momentum ${\cal F}_L$,
and energy ${\cal F}_E$ across the area
$z=0.5 Z_{max}$ as a function of time.
Time is
measured in units of
$t_{out} \approx 2200 t_i$,
where $t_i \equiv  2\pi r_i/{v_i}$,
where $r_i$ is the inner radius of the disk.}
\label{Figure 13}
\end{figure*}

  Another indication  of stationarity
is that the poloi-dal velocity becomes
parallel to the poloidal magnetic field.
    We observed, that the two vector
fields become parallel to a high accuracy
only after $t>t_{out}$.
  Figure 7a shows that the two vector
fields are close to being parallel even
at earlier times.

   One can get more complete
information about stationarity and validity
of the MHD flow  by comparing
the theory
reviewed in \S 3 with the simulation data.
 First,
the integrals of the motion, $\Lambda$,
$K$, $E$, $\Omega$, and $S$ (equations 4 - 8)
should be constants along
any magnetic field line.
   We  checked this by numerically
calculating these
integrals along the``reference''
magnetic field line.
  The calculated integrals
 are constants with  good
accuracy.   For example,
$|\Delta\Omega|/\Omega\lesssim0.06$ and
 $|\Delta S|/S\lesssim0.15$.
     Figure 14 shows variation of
the integrals.
     Note, that the
integrals are not strictly
constants in the region immediately
above the disk, because
 the grid is not fine enough in this
region due to the strong
gravitational force.
 Note that the integrals become constants
as a function of $\Psi$ much
later ($t \gtrsim t_{out}$)
than fluxes of mass, angular momentum,
and energy
become constants in time.

\placefigure{fig14}

\begin{figure*}[t]
\epsscale{0.5}
\plotone{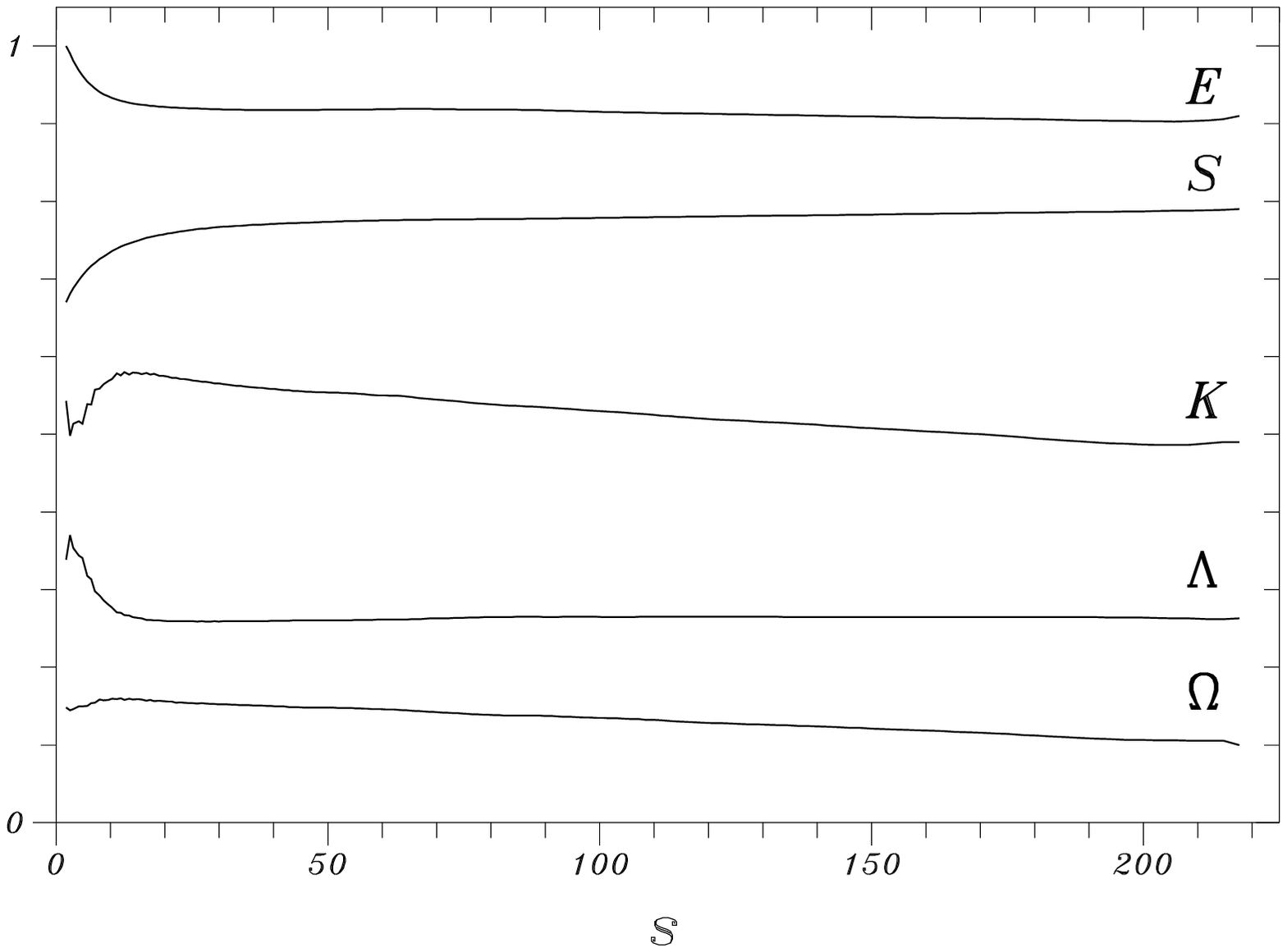}
\caption{
Numerically
calculated ``integrals of the motion'' (equations
1-5) as a function of distance $s$ along the
 ``reference'' magnetic
 field line.
In ideal MHD, the integrals should
be strictly constant.
In this plot the scale is such that the
maximum of the  $E$ integral is unity.
The other curves have been displaced
downward for clarity.
}
\label{Figure 14}
\end{figure*}

   Other comparisons of simulations with
theory have been done.  For example,
from the theory of stationary
flow it follows that
 fluxes of matter, angular momentum,
and energy flowing inside a
given flux tube
should be equal
to fluxes integrated
over the Alfv\'en surface,
equations (17)-(19).
We calculated these integrals
 in two ways, using
the data from our simulations.
Figures 15 a, b show these integrals as
 a function of $\Psi$. They almost coincide
in most of the region, excluding the region
of large values of $\Psi$.
   The latter
field lines have a
 high angle of inclination
relative to the axis and  do not pass
through the fast magnetosonic surface.
 These lines are marked by the
long-dashed lines on one of the
 curves, and by letter ``$f$''
on the $\Psi$ axis.

\placefigure{fig15}
\begin{figure*}[t]
\epsscale{0.5}
\plotone{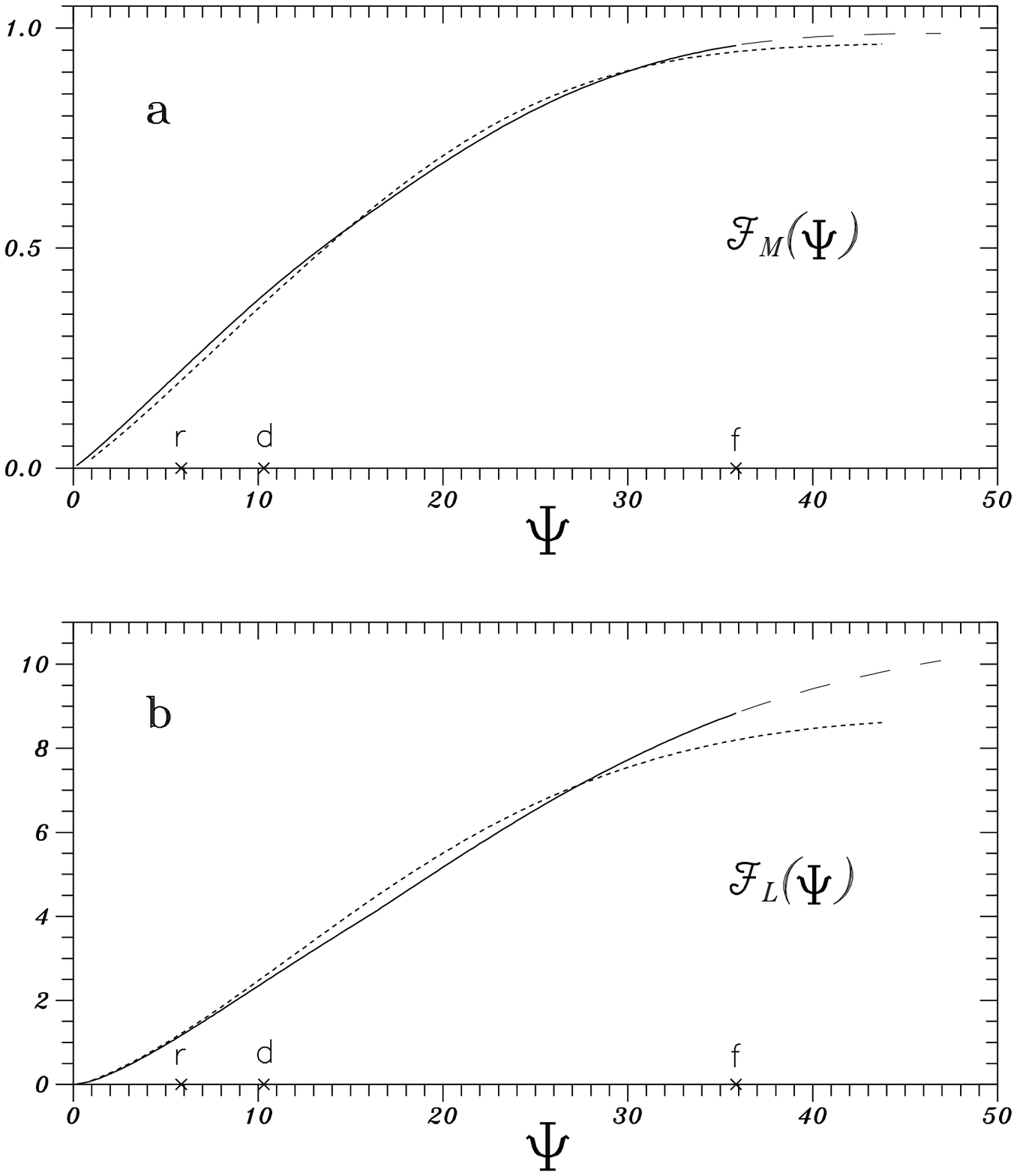}
\caption{
Panel (a) shows the matter flux
and panel (b) the angular momentum flux
as a function of $\Psi$
calculated in two ways.
  The solid line shows
the integrals calculated
using equations (11) and (12).
The dashed line shows
the integrals calculated
on the Alfv\'en surface
using equations (17) and (18).
Points ``$r$,'' ``$d$,''
and ``$f$'' on the  horizontal
axis are the same as
in Figure 12.
The long-dashed line shows the region
of strongly inclined field lines
which do not cross the fast magnetosonic surface.
These field lines are separated from the
other field lines by  ``$f$''
on the $\Psi$  axis.
}
\label{Figure 15}
\end{figure*}

Figure 16 shows the $\Psi$
dependence of the ratio of the
radii where a magnetic field
line crosses the Alfv\'en
surface and the disk,
  $\lambda=r_A(\Psi)/r_d(\Psi)$.
This ratio is of interest, because
the value of the angular
momentum per unit mass
carried by the outflowing matter
can be calculated  as
$$
\Lambda = \Omega ~r_A^2 =
\lambda^2 [GM r_d (\Psi)]^{1/2}~,
\eqno(42)
$$
from equation (14).
The fact that $\lambda$ is almost
constant
means that the specific
angular momentum
 is proportional
to $\sqrt{r_d}$ and can
be estimated in this way.
Specifically, $\Delta\lambda/\lambda
\lesssim 0.13$
for field lines which cross
the fast magnetosonic surface.

\placefigure{fig16}
\begin{figure*}[t]
\epsscale{0.5}
\plotone{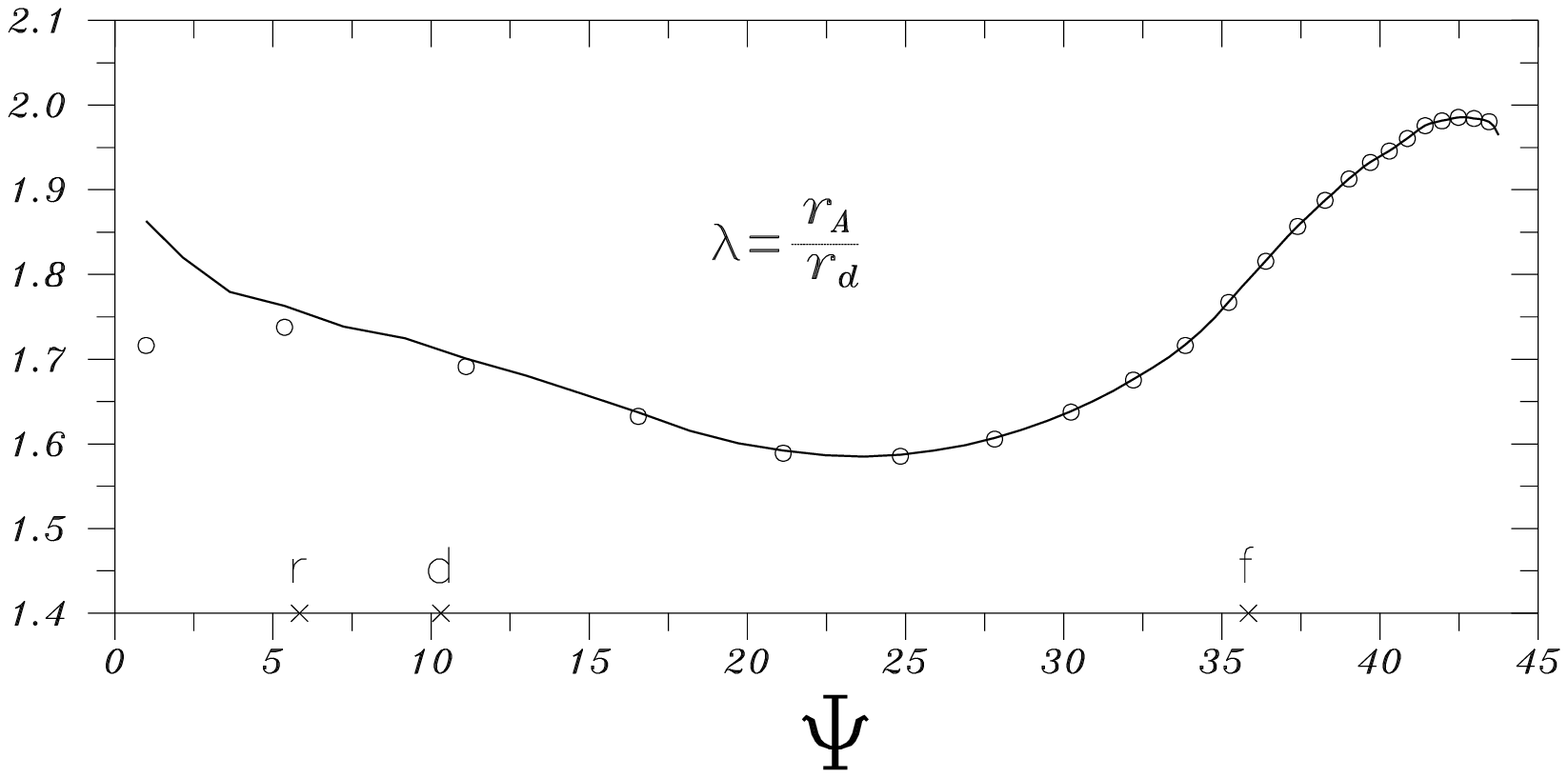}
\caption{
The ratio
$\lambda=r_A(\Psi)/r_d(\Psi)$
as a function of $\Psi$.
   Points ``$r$,'' ``$d$,'' and ``$f$''
are the same as in Figures 12.
}
\label{Figure 16}
\end{figure*}

    The fluxes of mass, energy,
and angular momentum
flowing out from the disk
depend of course on
the  magnetic field strength on the disk.
     Figure 17 shows the
dependence of the matter
outflow rate on the disk magnetic field.
    This dependence is analogous to that
derived by
Kudoh \& Shibata
(see Kudoh \& Shibata 1995, figure 2b,
and Kudoh \& Shibata 1997b,  Figure 24b)
who performed 1.5D analysis
of stationary MHD flows
at the fixed configuration of poloidal
magnetic field.

\placefigure{fig17}
\begin{figure*}[t]
\epsscale{0.5}
\plotone{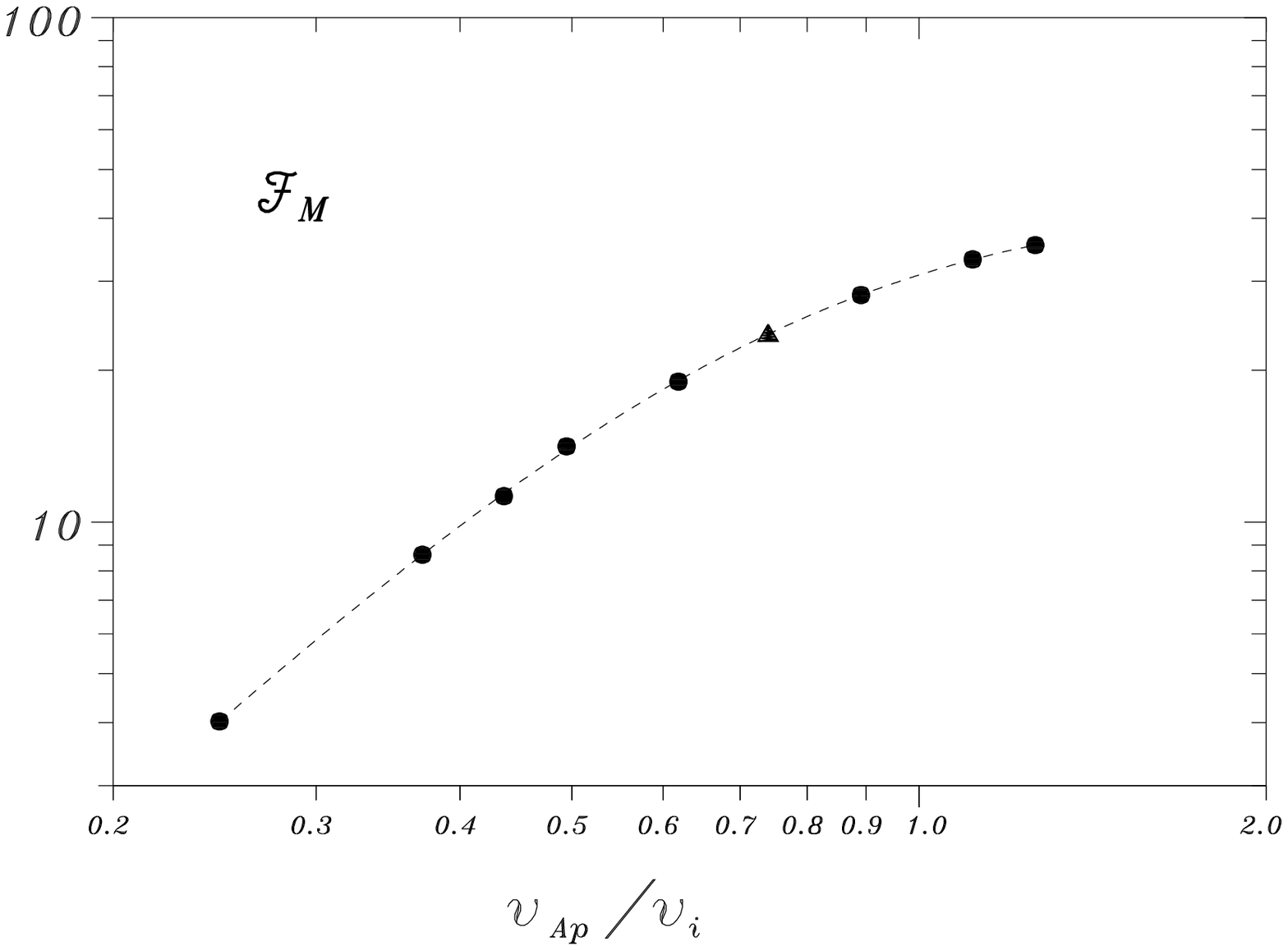}
\caption{
The dependence of matter flux from the disk
on the ratio $v_{Ap}/v_K$
evaluated at the point $r=r_i$.
Here, we changed only the magnitude of
the magnetic field $B_p$
while all other parameters were kept the same.
The triangle indicates the main case of the paper
while the circles were obtained from different
runs.
   The dashed line is simply a curve through
the points.
}
\label{Figure 17}
\end{figure*}

\subsection{Collimation}

   The  stationary MHD
outflows we find are approximately
spherical outflows with relatively small
 collimation within
the simulation region.
   Thus, the {\it collimation distance}
over which
the flow becomes collimated
(with divergence less
than, say, $10^o$) is much
larger than the size of
our simulation region.
   Figure 18 shows the
dependence of the angle
between the poloidal field direction
and the $z-$axis on the distance
along the ``reference'' magnetic
field line ($\theta_1$) and along the
``diagonal'' field line which
goes through the top right corner of
the simulation region
($\theta_2$).
   Both angles are relatively large ($> 30^o$)
near the disk and then
gradually decrease at larger
distances $s$ along the field line.
   This means that some
collimation occurs
near the disk but decreases
at larger distances.
   The angle $\theta_1$ for the
``reference'' field line changes
from $35^o$ at the disk
to $18^o$ at the top boundary where
the angle of the position vector
from the origin to
the $z-$axis is about $20.5^o$.
 The angle $\theta_2$ for
the ``diagonal'' field
changes from $49^o$ at
the disk  to $31^o$
at the top boundary where
the angle of the position
vector from the origin
 to the $z-$axis is about
$39^o$.

\placefigure{fig18}
\begin{figure*}[t]
\epsscale{0.5}
\plotone{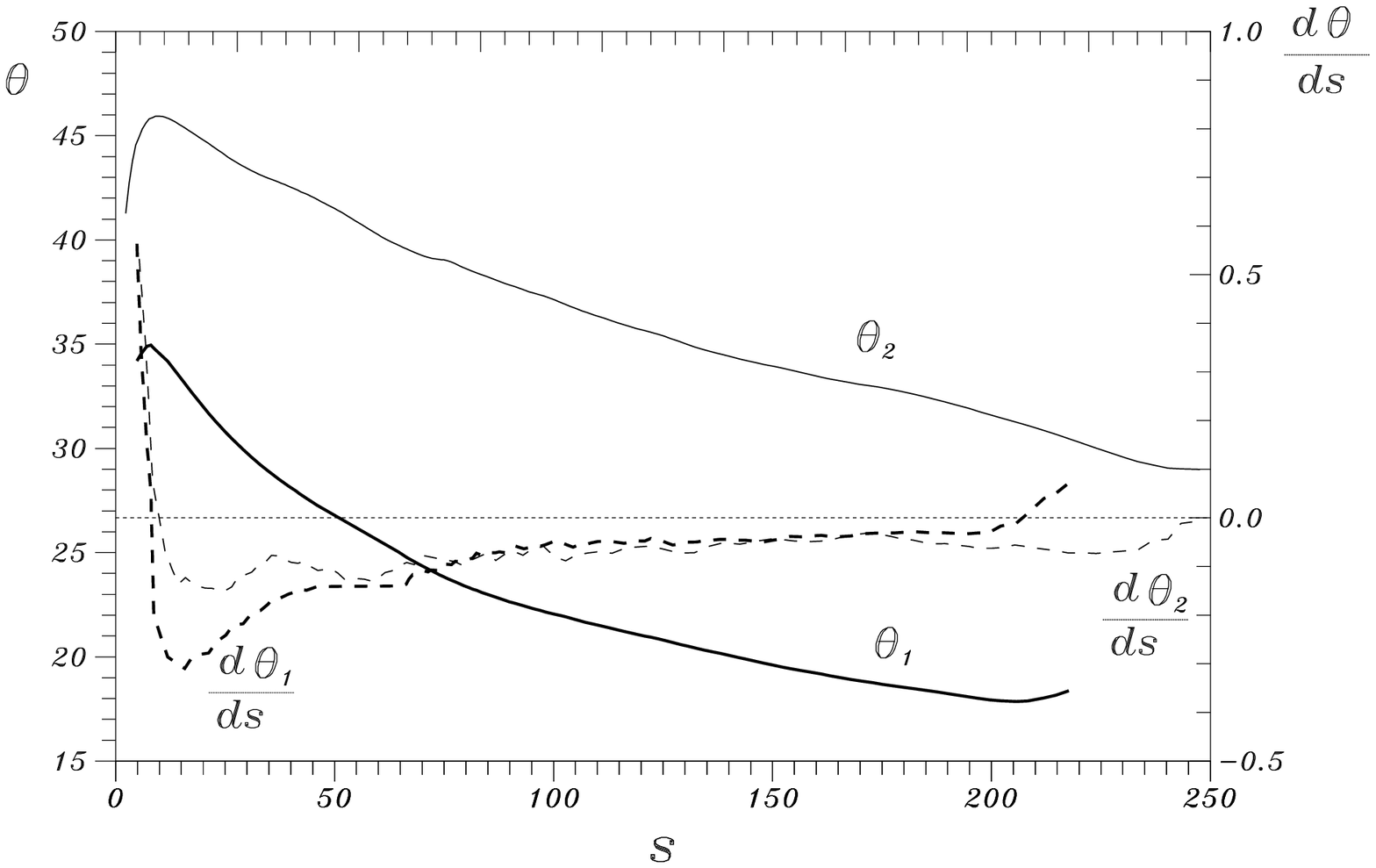}
\caption{
The figure shows
the dependence of the angle $\theta$ between
the poloidal magnetic
field and the $z-$axis
as a function of distance
$s$ along the field line.
The dependence of the derivatives
$\partial \theta/ \partial s$ is also shown.
The label ``$1$'' indicates
our ``reference'' magnetic
field line and  ``$2$''
the  ``diagonal'' field line discussed
in the text.
}
\label{Figure 18}
\end{figure*}

   An important question is whether the
outflow becomes
collimated at large distances
to form a cylindrical jet
parallel to the rotation axis,
or it continues as a wind
without collimation.
  To obtain information on this question,
we calculated the derivatives
$\partial \theta/ \partial s$
along magnetic field lines as also
shown in Figure 18.
  For both the
 ``reference''  and  the diagonal
field lines the derivatives
decrease and become very small at the
outer boundary.
   It appears that the derivatives
 continue to decrease, which
would mean that the collimation
decreases and goes to zero.
 (The turns at the ends
of the lines are connected with boundaries
and do not represent a
 real collimation effect.)
However, the present study does not
rule out the possible magnetic collimation of
the flow at much larger distances.
Separate simulations
in a much larger region
are needed to answer this important question.

  Earlier, Sakurai (1987) obtained stationary
flow solutions for
a split-monopole magnetic field and found
flows with very gradual collimation
at large distances from the origin.
  Our results are similar in this
respect to those of Sakurai.
However,
it is not clear that the flows will
magnetically collimate to cylinders
as predicted by Heyvaerts \& Norman (1989).
Simulations on a much larger region are needed
to answer this question.

  It is important to note
that analytic, self-similar solutions
for outflows
for cases of very gradually decreasing
poloidal magnetic field in the disk (unlike
the present split-monopole field)
 show  magnetic collimation
with increasing distance $z$ from the origin
(Contopoulos \& Lovelace 1994, Contopoulos 1995;
 Ostriker 1997).
  Further, note that outflows may be
collimated hydrodynamically
by the pressure of  surrounding, ambient
matter (Lovelace, Berk,
\& Contopoulos 1991; Frank \& Mellema 1996,
Mellema \& Frank 1998).
   This mechanism of collimation needs a separate
numerical investigation.

  It is of interest to
compare the present results
on collimation with those of our earlier studies,
Ustyugova et al. (1995)
and Romanova et al. (1997).
  Ustyugova et al. (1995) introduced
the treatment of the disk
as a boundary condition and found
non-stationary but
well-collimated outflows.
  Ustyugova et al. considered outflows
from a hot accretion disk  where the
sound speed $c_s\lesssim v_K$ and a weak
magnetic field $v_A^2<<v_K^2$
on the disk surface.
    For such conditions,
the main force driving
the outflow is the
matter pressure gradient while
the magnetic force is smaller.
    Also, Ustyugova et al. (1995) used
non-equilibrium initial conditions where
the rotation of the disk was started at $t=0$
with the corona of the disk not rotating.
    These conditions led to the
formation of a strong toroidal
magnetic field (by the winding up of the
poloidal field) and  a  strong
outward propagating
torsional Alfv\'en wave.
   The fact, that Alfv\'en
velocity was much smaller
than Keplerian veloicity allowed
the build up of the toroidal field
which in turn gave
strong collimation of the outflow.
   At later times in the simulation,
the twist of the field  relaxed, but the
matter pressure force
continued to push matter along the collimated
magnetic field lines.
   Thus, the Ustyugova et al. (1995) flows
are essentially different from the stationary
flows discussed in this paper where
the dominate driving forces are magnetic
and centrifugal  with the matter pressure
force negligible.

    Romanova et al. (1997) was the
first simulation study  to
obtain {\it stationary}
magneto-centrifugally  driven
outflows with relatively small matter pressure
force.
  The initial poloidal magnetic field
was a ``tapered'' monopole configuration.
  The outflows were found to be
uncollimated and are therefore similiar
to those discussed in this paper for the
split monopole initial field.

\subsection{ Illustrative Physical Values}

    Here, we discuss
physical values of parameters
for the case of MHD outflows
from the disk around a young star.
    The mass of the star is considered
to be $M=M_{\odot} \approx
2\times 10^{33} {\rm gm}$, and
the inner radius of the disk
is $r_i= 10^{11} {\rm cm}$, which
may be somewhat larger than the star's
radius.
  The magnetic
field threading the accretion
disk may arise from the ``shearing off''
and opening of the intrinsic stellar
field as discussed by Lovelace, Romanova,
\& Bisnovatyi-Kogan (1995).
   The reference field strength
at the inner edge of the
disk is taken to be
$B_i\equiv |{\bf B}_p(r_i,0)| =300 ~{\it G}$.
   The  computational region
extends from $(r,z)=0$ to
$(R_{max}, Z_{max})=(170r_i,200r_i)$
so that $R_{max} =1.7 \times 10^{13}~{\rm cm}$.
     The characteristic velocity
at the inner edge of the disk
is $v_i \equiv
(GM/r_i)^{1/2} \approx
365~{\rm km/s}$.   With our smoothed
potential the Keplerian velocity of the
disk at $r_i$ is $v_i/2^{3/4} \approx 0.6v_i$.
     The characteristic time at the inner
edge of the disk is
$t_i =2\pi r_i/v_{i} \approx 4.8~ {\rm hr}$.
     The density on the surface
of the disk  at $r=r_i$
can be written in terms of the Alfv\'en velocity as
$\rho_i=B_i^2/(4 \pi v_{Ap}^2) \approx
6.8\times10^{-12} (v_{i}/v_{Ap})^2 ~{\rm g/cm^3}$.

 We find that the mass outflow rate
from the top side of the disk
has the dependence
$$
\dot M_+ = {\cal F}_M r_i^2 B_i^2/v_{Ki}~,
$$
$$
\approx 4 \times 10^{-7} {M_{\odot} \over {\rm yr}}
{\cal F}_M
\left({r_i\over {10^{11}} {\rm cm}}\right)^{5\over2}
\left({B_i\over 300 {\rm G}}\right)^2
\left({M_{\odot} \over M}\right)^{1\over2}~,
\eqno(43)
$$
where ${\cal F}_M={\cal F}_M(v_A/v_i)$ is
the  stationary state value of ${\cal F}_M$
shown in  Figure 13 for our reference case.
   Figure 17 shows the dependence
of ${\cal F}_M$ on  $v_{Ap}/v_{i}$.
    For the reference
case, ${\cal F}_M\approx 5.6$ so that
$\dot M_+ \approx 2.2
\times 10^{-6}~{\rm M}_{\odot}/{\rm yr}$.
  The total mass outflow from the
disk is evidently $2 \dot M_+$.

Similarly, we can write the energy outflow rate
from the top of the disk as
$$
\dot E_+ = {\cal F}_E r_i^2 B_i^2 v_{Ki}~,
$$
$$
 \approx 1.5 \times 10^{34}
{{\rm erg} \over {\rm s}} {\cal F}_E
\left({r_i\over {10^{11}} {\rm cm}}\right)^{3/2}
\left({B_i\over
300 {\rm G}}\right)^2
\left({M \over M_\odot}\right)
^{1\over2}~.
\eqno(44)
$$
For given distribution of
velocities along the disk,
and given physical parameters at the
 inner edge of the disk,
we get ${\cal F}_E\approx 1.2$
and $\dot E_+ \approx 2.5 \times 10^{35}
{{\rm erg} / {\rm s}}$.
   The total energy outflow from the disk
is $2\dot{E}_+$.

Matter is accelerated from
 $v_p \approx 0.05 v_{i}\approx 18~ {\rm km/s}$
near the surface
of the disk  to $v_p \approx 0.47 v_{i} =
170~ {\rm km/s}$
at the outer boundary
of the simulation region (see Figure 10).

\section{Conclusions}

  We have studied MHD outflows
from a rotating, conducting
accretion disk using
axisymmetric simulations
The  disk was treated as a boundary
condition, and
the initial poloidal magnetic field
was taken to be a  split-monopole.
The main conclusions of this work are:

\begin{enumerate}

\item  In many different runs we
 observed the formation of
stationary MHD
outflows from the disk.
Close to the disk the main driving
force is the centrifugal force.
   At larger distances the main
driving force is the magnetic force
$\propto - {\bf \nabla}(rB_\phi)^2$.
The pressure gradient force is much
smaller than these forces and it
has no significant role in
driving the outflows.

\item  For the considered conditions,
the  slow magnetosonic surface
lies inside the disk.
  Above the
disk, the flow accelerates and
passes through the  Alfv\'en and fast
magnetosonic surfaces, which are almost parallel
to the disk.
   Within the simulation region, the outflow
accelerates from thermal velocity ($\sim c_s$)
to a much larger asymptotic poloidal flow
 velocity of the order of $0.5\sqrt{GM/r_i}$,
where $M$ is the mass of the central
object, and $r_i$ is the inner
radius of the disk.
  This asymptotic velocity is much larger than
the local escape speed and is larger than
fast magnetosonic speed by a factor of $\sim 1.75$.
  The {\it acceleration distance} for the outflow, over
which the flow accelerates from $\sim 0$
to say $90\%$ of the asymptotic speed, occurs
at a flow distance $\sim 80 r_i$.

\item
 The outflow is only
slightly collimated within the simulation region.
   The {\it collimation distance} for
the outflow, over which
the flow becomes collimated (with divergence less
than say $10^o$), is much larger than the size of
our simulation region.
  This ``poor'' collimation is similar
to that found in our earlier work
(Romanova et al. 1997) using a different
initial magnetic field and is
qualitatively similar
to the very gradual collimation found
by Sakurai (1987).
   MHD simulations using much larger
computatinal regions are needed to
determine the  collimation of
the outflow at large distances.
   Further, separate simulations are
also needed to study collimating influence
of an external medium
(Lovelace et al. 1991, Mellema \& Frank 1998).

\item  The stationarity of the MHD flows
was checked in a number of  ways,
including comparisons of simulation results
with predictions of theory of stationary
axisymmetric flows.
   We found that: (a) Fluxes of mass,
angular momentum, and energy
 across the surface
$z=0.5 Z_{max}$ become independent of
time with high
 precision at early times of simulations
$t<0.1 t_{out}$, where
$t_{out} \approx 2200t_i$ and
$t_i=2\pi r_i/\sqrt{GM/r_i}$.
(b) Integrals of the motion
become constants on flux surfaces with
accuracy $5\% - 15\%$
for $t\gtrsim t_{out}$.
(c) Vectors of
poloidal velocity are parallel to those of the
poloidal magnetic field
lines to a high accuracy.

\item  Different outer boundary
conditions on  the
toroidal magnetic field $B_\phi$
were investigated.
   We analyzed simulation results
using  and found that
collimation of the jet
and other characteristics of
the flow depend critically on the outer  boundary
 condition on $B_\phi$ (as well as the shape of the
simulation region as discussed below).
  We observed that the
outer ``free'' boundary condition on $B_\phi$
leads to an artificial force
which can give {\it apparent} magnetic
collimation of the flow.
     ``Force-free''
and ``force-balance'' outer boundary conditions
were also investigated.
  The ``force-free'' outer boundary
condition was found to give valid flow solutions
if the  simulation region is not narrow
in $r-$direction (compared with $z$-direction).

\item  The question of the
optimum shape of simulation
region was investigated. We have
shown that if region is narrow
in the $r-$direction,
then an essential part of the Mach cones
on the outer boundaries  may be directed towards
the inside of the computational region.
  This can lead to the  influence of the
boundary on the calculated
 flow and to artificial
collimation.
   This effect is reduced or absent if the
computational
region is approximately square,  if it is
elongated in the
 $r-$direction, or if it is spherical.  In
these cases the Mach cones tend to point
outside of the computaional region.

\end{enumerate}

\acknowledgements{
This work was supported in part by
NSF grant AST-9320068.
   The Russian
authors were supported in part
by RFFI Grant 96-02-17113.
   Also, the research described here
was made possible in part by
Grant No. RP1-173 of the U.S.
Civilian R\&D Foundation for the Independent
States of the Former Soviet Union.
   The work of
RVEL was also supported in part by NASA
grant NAG5 6311.}

\end{document}